\documentclass[prd,twocolumn,superscriptaddress,floatfix,preprintnumbers,nofootinbib,10pt,aps]{revtex4-2}

\usepackage[normalem]{ulem}
\usepackage{xcolor}
\usepackage{mathtools}
\usepackage{placeins}

\usepackage{enumitem}
\usepackage{mathrsfs}
\usepackage{physics}
\usepackage{amssymb,bm}
\usepackage{slashed}
\usepackage{graphicx}
\usepackage{tikz}
\usepackage{comment}
\usepackage{dsfont}
\usetikzlibrary{arrows.meta, decorations.pathmorphing, decorations.markings, decorations.pathreplacing, calc}

\definecolor{lcolor}{rgb}{0.5,0,0}
\definecolor{citcolor}{rgb}{0,0.3,0.0}
\usepackage{multirow}
\usepackage{booktabs}  

\newcommand{\der}{\mathrm{d}}
\newcommand{\eq}{Eq.\,}
\newcommand{\eqs}{Eqs.\,}
\newcommand{\fig}{Fig.\,}
\newcommand{\po}{\mathbb{P}}
\newcommand{\Nc}{N_\mathrm{c}}
\newcommand{\tmd}{_\text{TMD}}

\newcommand{\xpo}{x_\mathbb{P}}
\newcommand{\zp}{z_\gamma}
\newcommand{\as}{\alpha_\mathrm{s}}
\newcommand{\qs}{Q_\mathrm{s}}
\newcommand{\aem}{\alpha_{\mathrm{e.m.}}}

\newcommand{\bb}{\boldsymbol{\mathrm{b}}}
\newcommand{\rb}{\boldsymbol{\mathrm{r}}}
\newcommand{\sbo}{\boldsymbol{\mathrm{s}}}
\newcommand{\pbc}{\boldsymbol{\mathrm{p}}}
\newcommand{\ptc}{p_{\perp}}
\newcommand{\Kt}{K_\perp}
\newcommand{\PN}{P^-_N}
\newcommand{\bd}{b_{d\perp}}

\newcommand{\zq}{z}

\usepackage[breaklinks,colorlinks,citecolor=citcolor,urlcolor=blue,linkcolor=lcolor]{hyperref}

\begin{document}

\author{Paul Caucal}
\email{caucal@subatech.in2p3.fr}
\affiliation{SUBATECH UMR 6457 (IMT Atlantique, Universit\'e de Nantes, IN2P3/CNRS), 4 rue Alfred Kastler, 44307 Nantes, France}

\author{Patricia Gimeno-Estivill}
\email{patricia.p.gimenoestivill@jyu.fi}
\affiliation{Department of Physics, University of Jyväskylä,
P.O. Box 35, 40014 University of Jyväskylä, Finland}
\affiliation{Helsinki Institute of Physics,
P.O. Box 64, 00014 University of Helsinki, Finland}

\author{Tuomas Lappi}
\email{tuomas.v.v.lappi@jyu.fi}
\affiliation{Department of Physics, University of Jyväskylä,
P.O. Box 35, 40014 University of Jyväskylä, Finland}
\affiliation{Helsinki Institute of Physics,
P.O. Box 64, 00014 University of Helsinki, Finland}

\author{Farid Salazar}
\email{farid.salazar@temple.edu}
\affiliation{Department of Physics, Temple University, Philadelphia, Pennsylvania 19122, USA}
\affiliation{RIKEN-BNL Research Center, Brookhaven National Laboratory, Upton, New York 11973, USA}
\affiliation{Physics Department, Brookhaven National Laboratory, Upton, New York 11973, USA}

\title{Diffractive \texorpdfstring{$D^0$}{D0} photoproduction in ultraperipheral collisions at the LHC}

\begin{abstract}
We compute the cross section for coherent diffractive $D^0$ photoproduction in ultraperipheral lead--lead and
proton--lead collisions (UPCs) at the Large Hadron Collider (LHC) within the Color Glass Condensate (CGC) effective field theory. We include both the exclusive production of a charm-anticharm pair and diffractive production with an additional semi-hard
gluon, described in diffractive transverse-momentum-dependent (TMD) factorization. We show that the diffractive channel,
although suppressed by the strong coupling, dominates at large transverse momentum, where the exclusive channel is
suppressed by color transparency. Furthermore, we quantify nuclear effects at the parton level with a nuclear modification factor $R_{pA}$ at fixed photon energy, which shows a mild suppression for the diffractive channel and an enhancement growing with transverse momentum up to about a factor of 2 for the exclusive one.
We also use the diffractive cross section to correct our inclusive $D^0$ predictions to the CMS event selection, which
requires the target nucleus to break up, and compare them with the data. Finally, we provide predictions for inclusive
$D^0$ photoproduction in proton--lead UPCs.

\end{abstract}

\maketitle 
\tableofcontents

\section{Introduction}
In ultraperipheral collisions (UPCs), two nuclei pass each other without overlapping, and the target nucleus is probed by an electromagnetic field which acts as a flux of quasi-real photons 
\cite{Williams:1934ad,vonWeizsacker:1934nji,Baur:2001jj,Bertulani:2005ru,Klein:2019qfb}. Photoproduction in UPCs at the Large Hadron Collider (LHC) gives access to the gluon content of protons and nuclei at small longitudinal momentum fraction $x$, where gluon saturation, described by the Color Glass Condensate (CGC) effective field theory~\cite{Iancu:2003xm,Kovchegov:2012mbw,Morreale:2021pnn,Iancu:2026jlb}, is expected to become important. Photoproduction processes such as exclusive vector meson~\cite{STAR:2023nos,ALICE:2023jgu,CMS:2023snh,LHCb:2022ahs}, inclusive dijet~\cite{ATLAS:2024mvt} and open charm production~\cite{CMS:2025jjx} have recently been measured in ultraperipheral collisions.

In diffractive photoproduction, the photon scatters off the target through a color-singlet exchange referred to as the pomeron. This produces a large rapidity gap in the final state that can be measured experimentally. Since the exchange requires at least two gluons at the amplitude level \cite{Ryskin:1992ui}, diffractive cross sections depend more strongly on the gluon density of the target than inclusive ones, making them a good observable for the gluon saturation regime \cite{Kowalski:2008sa}. 

Coherent diffractive events, in which the target nucleus remains intact, are part of the inclusive photoproduction
cross section. Inclusive $D^0$ photoproduction in lead--lead UPCs has recently been computed in the CGC and collinear
frameworks~\cite{Gimeno-Estivill:2025rbw,Cacciari:2025tgr} and compared to CMS and ALICE
data~\cite{CMS:2025jjx,Nese:2025ohz}. These measurements select the $Xn0n$ neutron class, with no forward neutrons on
the photon-emitting side ($0n$) and at least one neutron emitted by the target ($Xn$). The $Xn$ condition removes the
coherent diffractive events, while the calculations include them, since they impose no condition on the target
($An0n$ class). A consistent comparison with the data therefore requires subtracting the coherent diffractive
contribution from the inclusive cross section, as was recently done in the  collinear factorization framework in Ref.~\cite{Guzey:2026lfb}.

In this paper, we present predictions for coherent diffractive $D^0$ photoproduction in lead--lead and proton--lead UPCs at LHC energies using the CGC framework. We include the exclusive production of a quark-antiquark pair and the production of a quark-antiquark-gluon system. In the presence of a hard scale $M_h\gg\qs$, the latter factorizes into hard factors and the diffractive transverse-momentum-dependent (TMD) gluon and quark distributions of the pomeron~\cite{Hauksson:2024bvv, Iancu:2022lcw,Caucal:2026xhj}, referred to as DTMDs. In the case of $D$-meson production in UPCs, the hard scale is provided by the transverse mass of the charm quark. Although formally suppressed by $\as$, the
quark-antiquark-gluon contribution dominates the diffractive cross section, by a large factor at large transverse
momentum, where the exclusive channel is suppressed by color transparency.

The experimental observable here is not a dijet measurement, but a single  inclusive cross section.
In a dijet measurement, both the hard relative momentum of the pair and its much smaller transverse momentum
imbalance are measured, and the ratio of these two scales generates large Sudakov logarithms. In the single-inclusive
cross section considered here, only the hard scale is measured. Consequently, we do not expect large Sudakov logarithms whose resummation would require TMD factorization. In
principle, the full diffractive cross section could then be computed at NLO in the CGC framework, once such a
calculation becomes available for massive quarks. 

However, the TMD factorization formula employed here provides the correct behavior when either the quark mass or the meson momentum is (much) larger than the saturation scale $Q_s$. In this regime, upon integrating over the semi-hard transverse momentum imbalance $\Kt$ of the quark-antiquark or quark-gluon pair to obtain the single-inclusive cross section, our approach effectively reduces to collinear factorization in the small-$x$ limit, formulated in terms of diffractive PDFs computed in the CGC picture. We expect these conditions to be met in most of the kinematic regime probed by the LHC measurements, although the
neglected corrections of order $\qs^2/M_h^2$, which are contained in a full NLO calculation, may be sizable in the
lowest $p_{D\perp}$ bins.

Furthermore, the TMD approach  has several practical  advantages. Firstly, it provides an approximation that is much easier to evaluate in the context of a detailed phenomenological study. Secondly, it expresses
the kinematics in terms of the target momentum fraction variables, which provides a more straightforward comparison to
the collinear diffractive parton distribution approach. It would be interesting to perform a quantitative study of the accuracy of the DTMD approach compared to the full CGC calculation in this kinematic regime, analogously to the work on diffractive DIS with massless quarks in Ref.\,\cite{Kaushik:2025roa}.

In addition to the diffractive cross section, which is the main result of this paper and a powerful probe of saturation physics in itself, we also present here an update to our predictions~\cite{Gimeno-Estivill:2025rbw} for inclusive $D$-meson production. We estimate the coherent diffractive contribution that is excluded by the $Xn0n$ selection in inclusive $D^0$ photoproduction, and compare the corrected inclusive cross section with the CMS data \cite{CMS:2025jjx}. Subtracting the diffractive contribution leads to a slight reduction in the $Xn0n$ cross section. Compared with Ref.~\cite{Gimeno-Estivill:2025rbw}, we now also include the DGLAP evolution of the fragmentation functions (FFs). Our predictions are consistent with the CMS data within uncertainties, although they tend to overestimate them at higher $p_{D\perp}$. Higher-order corrections to the inclusive cross section, and a treatment of the $D$-meson mass in the fragmentation beyond leading order in $m_D^2/p_{D\perp}^2$~\cite{Helenius:2018uul,Accardi:2009md}, could further improve the agreement. Finally, we present predictions for inclusive $D^0$ photoproduction in proton--lead UPCs, extending the lead--lead results of Ref.\,\cite{Gimeno-Estivill:2025rbw}.

The paper is organized as follows. Section~\ref{sec:pheno} describes the photon flux, the fragmentation functions and the dipole amplitude that we will use in our calculations. In Sec.~\ref{Sec: small-x}, we review the partonic cross sections for exclusive and TMD-factorized diffractive charm photoproduction, and evaluate  them numerically at the parton level. Section~\ref{sec:numerics} presents our numerical results for diffractive $D^0$ production in lead--lead and
proton--lead UPCs at the LHC, the corresponding nuclear modification factors, and the impact of the diffractive
contribution on the inclusive cross section in the $Xn0n$ neutron class, together with predictions for inclusive $D^0$
production in proton--lead UPCs. We summarize our results and discuss possible future developments in Sec.~\ref{sec: summary}. In Appendix~\ref{sec:flux_approximations}, we study how the transverse size of the target affects the photon flux, and compare our flux with the one used in the
STARlight event generator~\cite{Klein:2016yzr}.

\section{Ingredients for UPC phenomenology}
\label{sec:pheno}
Before discussing diffractive and inclusive $D^0$ photoproduction,
we begin by describing the photon flux, fragmentation functions and dipole amplitude used in the numerical analysis.

\subsection{Photon flux}
\label{sec:PhotonFlux}
Let us consider a relativistic nucleus $A$ emitting a flux of quasi-real photons that then interact with the target nucleus $B$, at nucleon-nucleon center-of-mass energy $\sqrt{s_{NN}}$. We denote by $\zp=\omega/E_A$ the ratio of the photon energy $\omega$ to the beam energy per nucleon, given in the nucleon-nucleon center-of-mass frame by $E_A=\sqrt{s_{NN}}/2$. In this manuscript, we use an effective photon flux $F^\text{eff}(\zp)$ that takes into account the transverse geometry of the colliding nuclei~\cite{Eskola:2024fhf}
\begin{multline}
\label{eq:eff_photon_flux}
   F^{\text{eff}}(\zp)=\int \der^2\rb\int\der^2\sbo \, f_{\gamma/A}^{\text{WS}}(\zp,r_\perp)\,\frac{T_B(\sbo)}{B}\\
   \times \Gamma_{AB}(|\rb-\sbo|)\,P_{\text{EMD}}^{\text{channel}}(|\rb-\sbo|) \,.   
\end{multline} 
Here $F^{\text{eff}}(\zp)$ is the number of photons per unit $\zp$ emitted by nucleus $A$. Our Pb--Pb cross sections are given for photons emitted by one of the two nuclei, with positive rapidity along the photon direction. The vector $\rb$ is the transverse position of the struck nucleon measured from the center of the photon-emitting $A$ nucleus, and $\sbo$ is its transverse position relative to the target $B$. Therefore, $b_\perp=|\rb-\sbo|$ is the transverse distance between the centers of the two nuclei, see Fig.~\ref{fig:UPC} in Appendix~\ref{sec:flux_approximations} for an illustration of the coordinates. In the following, we write $F(\zp)$ for $F^{\text{eff}}(\zp)$ in Pb--Pb collisions.  The bare photon flux $f_{\gamma/A}^{\text{WS}}(\zp,r_\perp)$ of an extended charge distribution is~\cite{Krauss:1997vr}
\begin{multline}
\label{eq:WS_bare_photon_flux}
f_{\gamma/A}^{\text{WS}}(\zp,r_\perp)
=
\frac{\aem Z^2}{\pi^2}
\frac{1}{\zp} \\ \times 
\Bigg|\int_0^\infty \mathrm{d}k_\perp\,
\frac{k_\perp^2\,F_A^{\text{WS}}\!\left(k_\perp^2+\zp^2m_N^2\right)}
{k_\perp^2+\zp^2m_N^2}
J_1\!\left(k_\perp r_\perp\right)\Bigg|^2 ,
\end{multline}
where $\aem=1/137$ is the fine-structure constant, $Z=82$ is the ion charge, $m_N=0.938$ GeV the nucleon mass and $J_1$ is the Bessel function of the first kind. The form factor $F_A^{\text{WS}}$ is taken as the Fourier transform of the Woods-Saxon (WS) distribution, normalized to $F_A^{\text{WS}}(0)=1$ \cite{Eskola:2024fhf}:
\begin{equation}\label{eq:WSFT}
    F_A^{\text{WS}}(q^2)=\frac{4\pi}{A}\int_0^\infty \der x \frac{x}{q}\sin(qx)\rho_A^{\text{WS}}(x) \,.
\end{equation}
 For the lead nucleus, we use the parameters $d = 6.49$ fm, $a = 0.54$ fm and $\rho_0 = 0.17$ fm$^{-3}$ in the Woods-Saxon distribution \cite{Eskola:2024fhf}
\begin{equation}
\label{eq:WS_density}
  \rho^{\text{WS}}_A(r) = \frac{\rho_0}{1+\exp\left(\dfrac{r-d}{a}\right)} \,,
\end{equation}
where $r=|\Vec{r}|$ is the radial distance from the center of the nucleus.

In order to have a UPC, the impact parameter must be large enough to prevent hadronic interactions between the nuclei,  $b_\perp \gtrsim 2R$.

Rather than as a sharp cutoff, we enforce this condition through the survival factor $\Gamma_{AB}(|\rb-\sbo|)$ in \eq\eqref{eq:eff_photon_flux}, which suppresses the flux at small impact parameters.
To be specific, it gives the probability of no hadronic interaction at impact parameter $b_\perp$ in the optical limit of the Glauber model \cite{Eskola:2024fhf}
\begin{equation}
\Gamma_{AB}(\bb) = \exp\big(-\sigma_{NN}T_{AB}(\bb)\big) \,.
\end{equation}
Here $\sigma_{NN} = 92$~mb \cite{Guzey:2026lfb,ParticleDataGroup:2020ssz} is the total\footnote{Unlike in many areas of heavy ion physics, the total cross section is used instead of the inelastic cross section. This is because elastic nucleon-nucleon collisions also transfer a transverse momentum to the nucleon leading to nuclear breakup, which violates the UPC condition that the nucleus remain intact~\cite{Eskola:2024fhf}.} nucleon-nucleon cross section at $\sqrt{s_{NN}}=5.36$ TeV and $T_{AB}$ the nuclear overlap function 
\begin{equation}
T_{AB}(\bb) = \int \der^2\sbo \,T_A(\sbo)\,T_B(\bb - \sbo) \,.
\end{equation}
The nuclear thickness functions $T_{A,B}$ are the Woods-Saxon density \eqref{eq:WS_density} integrated along the beam axis $z$:
\begin{equation}
\label{eq:thickness}
T_{A,B}(\sbo) = \int_{-\infty}^{\infty} \der z\,\rho^{\text{WS}}_{A,B}\left(\sqrt{z^2+s_\perp^2}\right) \,,
\end{equation}
normalized as $\int \der^2\sbo\, T_{B,A}(\sbo)=B,A$.\footnote{
Note that the form factor in \eq\eqref{eq:WSFT} is a three-dimensional Fourier-transform. Thus the integral over $z$, i.e. the thickness function, is related to the form factor evaluated at $k^z=0$.}
Hence, $T_B(\sbo)/B$ is the probability density for the transverse position of the struck nucleon in the target. 
Equation~\eqref{eq:eff_photon_flux} is therefore the photon flux averaged over the nucleon distribution of $B$. The remaining integral over the transverse position $\rb$ is equivalent to an integral over the impact parameter $\bb$.

Finally, the electromagnetic-dissociation (EMD) factor $P_{\text{EMD}}^{\text{channel}}(b_\perp)$ in \eq\eqref{eq:eff_photon_flux} encodes the neutron-tagging requirement measured experimentally by the zero-degree calorimeter (ZDC),
\begin{equation}
\label{eq:channels}
P_{\text{EMD}}^{\text{channel}}(b_\perp) =
\begin{cases}
1 & AnAn \\[2pt]
P(b_\perp) & An0n\\[2pt]
P(b_\perp)P(b_\perp) & 0n0n\\[2pt]
P(b_\perp)[1-P(b_\perp)] & Xn0n
\end{cases}\,,
\end{equation}
where $P(b_\perp)$ is the probability of no electromagnetic dissociation. In each channel label, the first entry refers to the neutrons detected from the target nucleus and the second to those from the photon-emitting nucleus. Here $An$ denotes no requirement on the number of neutrons, $0n$ requires zero neutrons, and $Xn$ requires at least one neutron.

In the predictions for coherent diffraction presented below, we impose the $An0n$ requirement using the parametrization~\cite{Baur:1998ay}
\begin{equation}
\label{eq:EMD}
P(b_\perp) = \exp\left(-\frac{S}{b_\perp^2}\right)\,.
\end{equation}
\begin{figure*}
    \centering
    \includegraphics[width=0.9\linewidth]{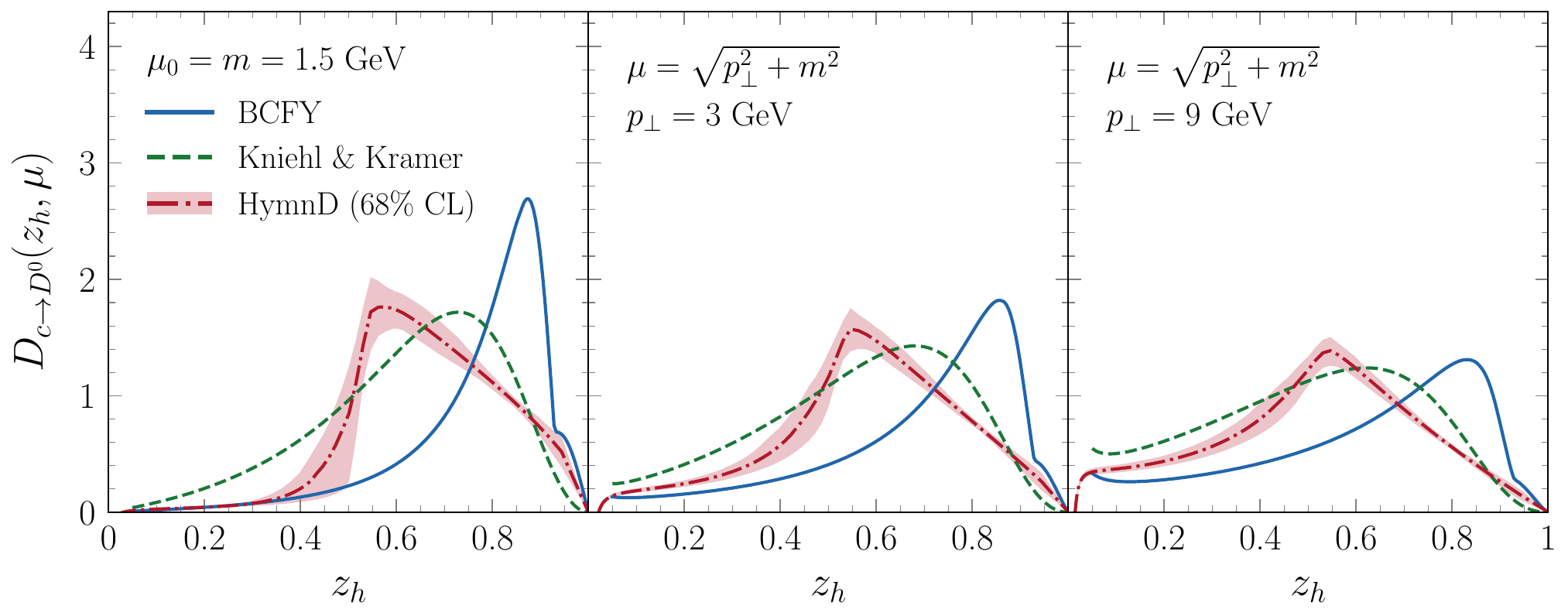}
    \caption{$c\to D^0$ fragmentation functions $D(z_h,\mu)$ as a function of $z_h$ for the BCFY
    (solid), KK (dashed) and HymnD (dash-dotted) parametrizations at the initial
    scale $\mu_0=m=1.5$~GeV (left) and at $\mu=m_T=\sqrt{m^2+p_{\perp}^2}$ after DGLAP evolution (middle and right).
    The band shows the 68\% confidence level (CL) of the HymnD FF.}
    \label{fig:FF}
\end{figure*}
We use the value $S=(17.4~\text{fm})^2$~\cite{Baltz:1997di,Baur:1998ay}, as in Ref.~\cite{Guzey:2026lfb}, which includes all electromagnetic excitations of the nucleus, such as quasideuteron absorption and nucleon resonances. Another value in the literature is $S=(10.4~\text{fm})^2$~\cite{Bertulani:1987tz,Baur:1998ay}, which considers only the giant dipole resonance.

In Appendix~\ref{sec:flux_approximations} we compare our effective flux with the STARlight~\cite{Klein:2016yzr} parametrization of Ref.\,\cite{Eskola:2024fhf}, and the effect on the cross sections is quantified in Sec.~\ref{sec:numerics}. In the Appendix we also compare the two EMD parametrizations and show how \eq\eqref{eq:eff_photon_flux} simplifies when the bare photon flux \eq\eqref{eq:WS_bare_photon_flux} is evaluated at the impact parameter $b_\perp$, rather than at the nucleon position $r_\perp$.

In proton--lead UPCs, the photon flux is dominated by the lead nucleus due to the $Z^2$ enhancement in \eq\eqref{eq:WS_bare_photon_flux}, such that the proton acts as the hadronic target. In this configuration, the photon flux emitted by the nucleus is given by
\begin{equation}
\label{eq:pAphotonFlux}
        F(\zp)=\int \der^{2}\bb\,f_{\gamma/A}^{\text{WS}}(\zp,b_\perp)\,\Gamma_{pA}(\bb) \,,
\end{equation}
where $f_{\gamma/A}^{\text{WS}}$ is the bare flux \eqref{eq:WS_bare_photon_flux} evaluated at the center of the proton, and $\Gamma_{pA}$ accounts for the suppression of hadronic interactions. This is given in the optical limit of the Glauber model as
\begin{equation}
\label{eq:pA_survival_factor}
    \Gamma_{pA}(\bb)=\exp(-\sigma_{NN}T_A(\bb)) \,,
\end{equation}
where $\sigma_{NN}=99$ mb is the total nucleon-nucleon cross section for $\sqrt{s_{NN}}=8.16$ TeV \cite{Guzey:2026lfb, ParticleDataGroup:2020ssz}. A comparison of \eq\eqref{eq:pAphotonFlux} with a
point-like flux with a sharp cutoff is shown in Appendix~\ref{sec:flux_approximations}.

To compute the differential $D^0$ cross section, we convolve the partonic cross section with the photon flux by integrating over the photon energy fraction $\zp$. The integral has a lower limit set by requiring enough energy to produce a charm quark with momentum\footnote{We use the convention $x^{\pm}=(x^0\pm x^3)/\sqrt{2}$ for the light-cone coordinates.} $p^+$, $q^+\geq p^+$. With $q^+ = \sqrt{2}\zp E_A$, the minimum value is
\begin{equation}
\label{eq:qmin}
\zp^{\min} = \frac{q^+_{\min}}{\sqrt{2}E_A} = \frac{p^+}{\sqrt{2}E_A} = \frac{m_T e^y}{2E_A} \,,
\end{equation}
where $m_T=\sqrt{m^2+p_\perp^2}$ is the transverse mass of the produced charm quark, with $m=1.5$~GeV the charm quark mass, $p_\perp$ its transverse momentum and $y$ its rapidity.
\subsection{Fragmentation functions}
\label{sec:FF}
The charm quark produced in the hard scattering hadronizes into a $D^0$ meson. This is described by the fragmentation function $D(z_h,\mu)$, where $z_h=p_{D\perp}/\ptc$ is the fraction of the
charm-quark transverse momentum carried by the meson\footnote{We assume that the $D^0$ has the same rapidity as the charm quark, $y_D=y_c$, which is a standard choice in collinear
factorization~\cite{Cacciari:1998it,Cacciari:2025tgr}. However, this choice is not unique, since for a massive quark fragmenting into a hadron of a different mass, the four-momenta cannot be collinear, and the mapping between quark and hadron kinematics depends on the definition of $z_h$. For example, defining it as the energy fraction in the collision center-of-mass frame~\cite{Helenius:2018uul}, $z_h = E_D/E_c = m^D_T \cosh(y_D)/(m_T \cosh(y_c))$, as is usual in $e^+e^-$ annihilation, with the hadron three-momentum parallel to that of the quark, would lead to $|y_D|\leq|y_c|$. Using the photon-direction
light-cone momentum fraction, as is usual in SIDIS, would give a shift
$y_D-y_c =\ln(z_h m_T/m^D_T)$, with $m^D_T=\sqrt{m_D^2+p_{D\perp}^2}$. These prescriptions differ by corrections
of order $m_D^2/p_{D\perp}^2$, which are beyond the accuracy of leading-power
fragmentation, and they all reduce to $y_D=y_c$ at high momentum, where the masses can be
neglected. The choice made here might have a sizeable effect on our results in the low $p_{D\perp}$ region.}, with $0<z_h<1$.

In our numerical analysis, we use three different parametrizations of the FFs:
\begin{enumerate}[leftmargin=1em]
    \item Heavy Meson collaboration (HymnD) \cite{Epele:2026dia}. The FF for prompt $D^0$ is extracted
    from an NLO fit to experimental data from LEP and the LHC. Its uncertainty is quantified within a Monte
    Carlo framework \cite{Costantini:2024wby} by fitting independent data replicas.

    \item Braaten-Cheung-Fleming-Yuan (BCFY) \cite{Braaten:1994bz}. This model is a linear combination of a pseudoscalar function $D_P(z_h,r)$,
    corresponding to direct hadronization $c\rightarrow D^0$, and a vector function $D_V(z_h,r)$
    describing the indirect contribution $c\rightarrow D^*\rightarrow D^0$. Note that the functions $D_P$ and $D_V$ given in Ref.\,\cite{Braaten:1994bz} are not normalized. We therefore normalize each of them separately to unity, so that the FF integrated over $z_h$ gives the total $c\to D^0$ fragmentation fraction.
    We use $r=0.1$ for the non-perturbative parameter appearing in the parametrization, following Refs.\,\cite{Cacciari:2003zu,Guzey:2026lfb}.    
    \item Kniehl-Kramer (KK) \cite{Kniehl:2006mw}. This has the form
    \begin{equation}
        D(z_h)=N\frac{z_h(1-z_h)^2}{[(1-z_h)^2+\epsilon z_h]^2} \,,
    \end{equation}
    with parameters $N=0.694$ and $\epsilon=0.101$, at the initial scale $\mu_0=m$, obtained from a leading-order fit to LEP1 data.
\end{enumerate}

We evolve the BCFY and KK FFs from the initial scale $\mu_0=m=1.5$~GeV to the factorization scale
$\mu_F$ using leading-order time-like DGLAP evolution with the EKO package~\cite{Candido:2022tld}. The HymnD FF is taken directly at $\mu_F$ from its next-to-leading-order evolution \cite{Epele:2026dia}. Although our partonic cross sections are computed at leading order, we include this NLO set as an independent estimate of the FF uncertainty.

For a given FF, the dominant source of
uncertainty in our numerical predictions is the variation of the factorization scale $\mu_F$ of the
FFs and of the renormalization scale $\mu_R$ of $\as$ in \eq\eqref{eq:hard factor}. We estimate
it with a seven-point variation around the central scale $\mu=m_T=(m^2+(p_{D\perp}/z_h)^2)^{1/2}$, i.e.
$\mu_F,\mu_R\in\{0.5\mu,\,\mu,\,2\mu\}$ with $1/2\leq\mu_F/\mu_R\leq2$. We set $\mu_F=\mu_0$ when $\mu_F<\mu_0$.

The three FFs are compared in \fig\ref{fig:FF}. At the initial scale, the BCFY function peaks at large $z_h\approx0.87$, while the KK and HymnD functions peak at smaller $z_h$. The evolution lowers and broadens all three distributions and slightly shifts their peaks to smaller $z_h$, but BCFY remains the hardest.
Since the charm spectrum falls steeply with transverse momentum, the $D^0$ cross section at large values of $p_{D\perp}$ is dominated by the large-$z_h$ region of the FF. Thus the BCFY parametrization leads to the largest cross sections, except at the lowest $p_{D\perp}$, where the spectrum is flatter and the three FFs give similar results.
\subsection{Dipole amplitude}
\label{sec:dipole}
For the numerical study of $D^0$ production in UPCs, we use the dipole amplitude 
evolved with the running-coupling Balitsky-Kovchegov (rcBK) equation
\cite{Balitsky:1995ub,Kovchegov:1999yj,Balitsky:2006wa} up to an evolution rapidity given by $Y_{\mathbb{P}}=\ln x_0/\xpo$. The BK evolution is taken to start from a nonperturbative initial condition at
$x_0=0.01$, and we freeze the dipole amplitude to this initial condition for $\xpo>x_0$. Inspired by the McLerran-Venugopalan (MV) model
\cite{McLerran:1993ka,McLerran:1993ni,McLerran:1994vd}, the initial condition for a proton target is parametrized as
\begin{equation}
\label{eq:ICBK_proton}
    \mathcal{N}_p(r_\perp)=1-\exp\bigg[-\frac{(r_\perp^2Q^2_{s0})^\gamma}{4}
    \ln\bigg(\frac{1}{r_\perp\Lambda_{\text{QCD}}}+e_c\cdot e\bigg)\bigg]\,.
\end{equation}
Here we use the MV$^e$ fit of Ref.\,\cite{Lappi:2013zma}, where $\gamma=1$ is fixed, and the parameters
$Q_{s0}^2=0.06$~GeV$^2$, $e_c=18.9$ and the proton transverse area
$S_\perp=\sigma_0/2=16.36$~mb are determined by a fit to the HERA inclusive DIS data
\cite{H1:2009pze}. The running coupling is evaluated in coordinate space with the Balitsky prescription~\cite{Balitsky:2006wa}, with $C^2=7.2$, $\Lambda_{\rm QCD}=0.241$~GeV and $\as$ frozen at $0.7$ in the infrared, as in~\cite{Lappi:2013zma}.

For a nuclear target, the rcBK equation is solved independently at each impact parameter $\bd$, which is the distance of the dipole center relative to the center of the target. The $\bd$ dependence of the initial condition is determined within the optical
Glauber model:
\begin{multline}
\label{eq:ICBK_nucleus}
    \mathcal{N}_A(r_\perp,\bd)=1-\exp\biggl[-T_A(\bd)\,S_\perp\,
    \frac{(r_\perp^2 Q_{s0}^2)^\gamma}{4}\\
    \times\ln\biggl(\frac{1}{r_\perp\Lambda_{\text{QCD}}}+e_c\cdot e\biggr)\biggr]\,.
\end{multline}
Here $T_A(\bd)$ is the nuclear thickness function of \eq\eqref{eq:thickness} normalized to $A$, which determines the transverse geometry of the nucleus. The dipole amplitude of the nucleus is computed up to $\bd= 6.3$~fm, where the nuclear saturation scale given by the optical Glauber model is the same as for a proton. Beyond it, we use the dilute limit $\mathcal{N}_A \approx T_A(\bd)\,S_\perp\,\mathcal{N}_p$, which adds a contribution proportional to the proton-target cross section \cite{Lappi:2013zma}. 
Equation \eqref{eq:ICBK_nucleus} introduces no parameters fitted to nuclear data beyond the standard Woods-Saxon parametrization, so the nuclear modification of the cross sections is a prediction of the framework.

 We consider cross sections integrated over the momentum transfer $t$, since $t$ cannot realistically be measured in coherent nuclear diffraction at the LHC. For both the exclusive and the diffractive TMD contributions, the integration over $t$ sets the amplitude and its complex conjugate to have the same impact parameter, so that, neglecting the minimal momentum transfer $|t_{\min}|\approx(\xpo m_N)^2$ (see Sec.~\ref{sec:inclusive}), the cross sections are proportional to the transverse area $S_\perp$ of a homogeneous target. With the $\bd$ dependence of the nuclear dipole amplitude in \eq\eqref{eq:ICBK_nucleus}, this area factor (see e.g.\ \eqs\eqref{eq:ExclusiveCrosssection}, \eqref{eq:GDTMD} and \eqref{eq:qDTMD}) is replaced by an integral over $\bd$,
\begin{equation}
    S_\perp\,\big[\mathcal{F}\big]^2\to\int\der^2\bd\,\big[\mathcal{F}(\bd)\big]^2\,.
\end{equation}

To estimate the uncertainty of the nuclear modification related to the dipole amplitude, we use $100$ posterior samples of the parameters $Q_{s0}^2$, $e_c$ and $\sigma_0/2$, together with the parameter of the running coupling, from the Bayesian
fit to HERA data of Ref.\,\cite{Casuga:2023dcf}. For each sample, the proton and nuclear amplitudes are constructed with the same parameters, and $R_{pA}$ is computed with the same sample. These samples come from a different fit than our central dipole amplitude, so we use only their relative spread. The band labeled ``dipole fit'' in the figures is the standard deviation of $R_{pA}$ over the samples, divided by their mean and applied around the central result. Because the proton and nuclear amplitudes vary together, this uncertainty largely cancels in the ratio.

\section{Coherent diffraction in the dipole picture}
\label{Sec: small-x}
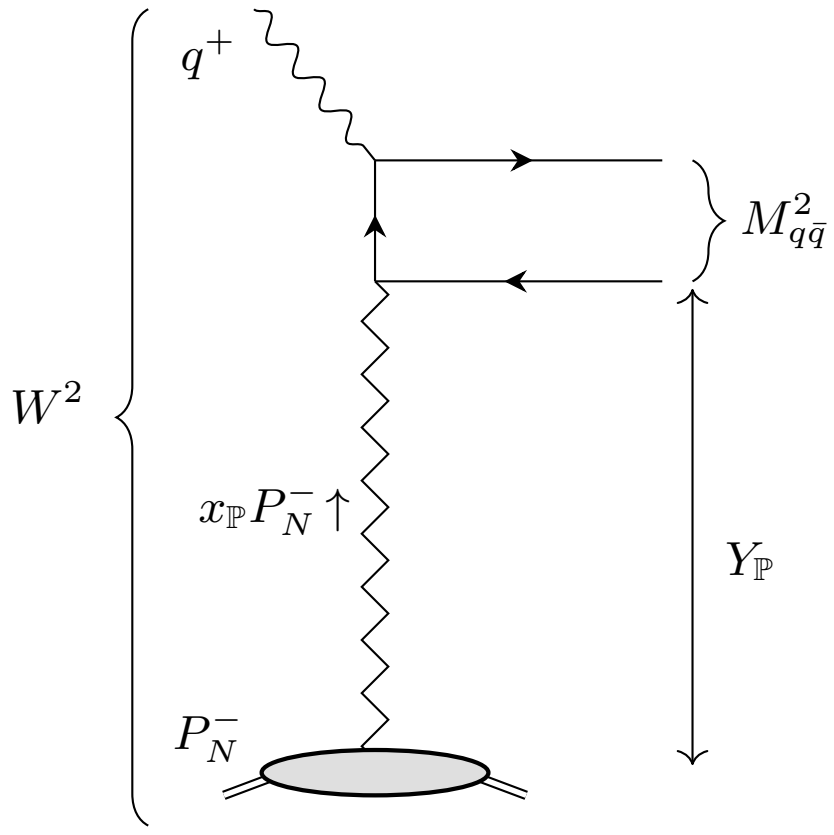
\begin{figure}[!t]
    \centering
    \begin{tikzpicture}[
  fermion/.style={postaction={decorate},
    decoration={markings, mark=at position 0.55 with {\arrow{Stealth[length=1.5mm, width=1.5mm, inset=0.5mm]}}}},
  photon/.style={decorate, decoration={snake, amplitude=2pt, segment length=8pt}},
  pomeron/.style={decorate, decoration={zigzag, amplitude=2.5pt, segment length=10pt}},
  brace/.style={decorate, decoration={brace, amplitude=6pt}}
]
  \coordinate (v1) at (0,0);
  \coordinate (v2) at (0,-0.8);
  \coordinate (b)  at (0,-4.05);   
  \draw[photon] (-0.8,1) -- (v1);
  \node at (-1.1,0.7) {$q^+$};
  \draw[fermion] (v1) -- (1.9,0);
  \draw[fermion] (1.9,-0.8) -- (v2);
  \draw[fermion] (v2) -- (v1);
  \draw[brace] (2.1,0.0) -- (2.1,-0.8);
  \node[right] at (2.3,-0.4) {$M_{q\bar q}^2$};
  \draw[pomeron] (v2) -- (0,-4);
  \node[left] at (-0.05,-2.3) {$x_{\mathbb{P}} P_N^- \!\uparrow$};
  \draw[double, double distance=1pt] (-1,-4.2) -- (-0.6,-4.05);
  \draw[double, double distance=1pt] (0.6,-4.05) -- (1,-4.2);
  \draw[thick, fill=gray!25] (b) ellipse (0.75 and 0.15);
  \node at (-1.1,-3.8) {$P_N^-$};
  \draw[{To[length=1mm]}-{To[length=1mm]}] (2.1,-0.85) -- (2.1,-4);
  \node[right] at (2.2,-2.65) {$Y_{\mathbb{P}}$};
  \draw[brace] (-1.5,-4.4) -- (-1.5,1);
  \node[left] at (-1.8,-1.6) {$W^2$};
\end{tikzpicture}
    \caption{Coherent exclusive photoproduction of a charm quark pair in the dipole picture. The zigzag line represents the exchange of a color-neutral pomeron, experimentally observed as a rapidity gap $Y_{\mathbb{P}}$. The momenta and diffractive variables are defined in the main text.
    }
    \label{fig:exclusive}
\end{figure}

\begin{figure*}[t]
    \centering
\begin{tikzpicture}[
  fermion/.style={postaction={decorate},
    decoration={markings, mark=at position 0.55 with {\arrow{Stealth[length=1.5mm, width=1.5mm, inset=0.5mm]}}}},
  photon/.style={decorate, decoration={snake, amplitude=2pt, segment length=8pt}},
  gluon/.style={decorate, decoration={coil, amplitude=2pt, segment length=3.8pt, aspect=0.8}},
  pomeron/.style={decorate, decoration={zigzag, amplitude=2.5pt, segment length=10pt}},
  brace/.style={decorate, decoration={brace, amplitude=6pt}}
]
  \coordinate (v1) at (0,0);
  \coordinate (v2) at (0,-0.6);
  \coordinate (g)  at (0,-1.6);
  \coordinate (b)  at (0,-4.85);
  \draw[photon] (-0.6,0.8) -- (v1);
  \node at (-0.9,0.7) {$q^+$};
  \draw[fermion] (v1) -- (1.6,0);
  \draw[fermion] (1.6,-0.6) -- (v2);
  \draw[fermion] (v2) -- (v1);
  \draw[brace] (1.8,0.0) -- (1.8,-0.6);
  \node[right] at (2,-0.3) {$M_{q\bar q}^2
  $};
  \draw[gluon] (g) -- (v2);
  \node[left] at (-0.1,-1.2) {$x_g x_{\mathbb{P}} P_N^- \!\uparrow$};
  \draw[gluon] (g) -- (1.6,-1.6);
  \node[above] at (1.2,-1.6) {$(1-x_g) x_{\mathbb{P}} P_N^-$};
  \draw[pomeron] (g) -- (0,-4.8);
  \node[left] at (-0.05,-3.2) {$x_{\mathbb{P}} P_N^- \!\uparrow$};
  \fill (g) circle (2.2pt);
  \draw[double, double distance=1pt] (-1,-5.0) -- (-0.6,-4.85);
  \draw[double, double distance=1pt] (0.6,-4.85) -- (1,-5.0);
  \draw[thick, fill=gray!25] (b) ellipse (0.75 and 0.15);
  \node at (-1.25,-4.8) {$P_N^-$};
  \draw[{To[length=1mm]}-{To[length=1mm]}] (2.1,-1.7) -- (2.1,-4.8);
  \node[right] at (2.2,-3.25) {$Y_{\mathbb{P}}$};
  \draw[brace] (-1.6,-5.1) -- (-1.6,0.85);
  \node[left] at (-1.8,-2.1) {$W^2$};
  \node at (-2.4,0.8) {(a)};
\end{tikzpicture}
\hspace{1.5cm}
\begin{tikzpicture}[
  fermion/.style={postaction={decorate},
    decoration={markings, mark=at position 0.55 with {\arrow{Stealth[length=1.5mm, width=1.5mm, inset=0.5mm]}}}},
  photon/.style={decorate, decoration={snake, amplitude=2pt, segment length=8pt}},
  gluon/.style={decorate, decoration={coil, amplitude=2pt, segment length=3.8pt, aspect=0.8}},
  pomeron/.style={decorate, decoration={zigzag, amplitude=2.5pt, segment length=10pt}},
  brace/.style={decorate, decoration={brace, amplitude=6pt}}
]
  \coordinate (v1) at (0,0);
  \coordinate (v2) at (0,-0.6);
  \coordinate (g)  at (0,-1.6);
  \coordinate (b)  at (0,-4.85);
  \draw[photon] (-0.6,0.8) -- (v1);
  \node at (-0.9,0.7) {$q^+$};
  \draw[fermion] (v1) -- (1.6,0);
  \draw[gluon] (v2) -- (1.6,-0.6);
  \draw[fermion] (v2) -- (v1);
  \draw[brace] (1.8,0.0) -- (1.8,-0.6);
  \node[right] at (2,-0.3) {$M_{qg}^2$};
  \draw[fermion] (g) -- (v2);
  \node[left] at (-0.1,-1.2) {$x_q x_{\mathbb{P}} P_N^- \!\uparrow$};
  \draw[fermion] (1.6,-1.6) -- (g);
  \node[above] at (1.2,-1.6) {$(1-x_q) x_{\mathbb{P}} P_N^-$};
  \draw[pomeron] (g) -- (0,-4.8);
  \node[left] at (-0.05,-3.2) {$x_{\mathbb{P}} P_N^- \!\uparrow$};
  \fill (g) circle (2.2pt);
  \draw[double, double distance=1pt] (-1,-5.0) -- (-0.6,-4.85);
  \draw[double, double distance=1pt] (0.6,-4.85) -- (1,-5.0);
  \draw[thick, fill=gray!25] (b) ellipse (0.75 and 0.15);
  \node at (-1.25,-4.8) {$P_N^-$};
  \draw[{To[length=1mm]}-{To[length=1mm]}] (2.1,-1.7) -- (2.1,-4.8);
  \node[right] at (2.2,-3.25) {$Y_{\mathbb{P}}$};
  \draw[brace] (-1.6,-5.1) -- (-1.6,0.85);
  \node[left] at (-1.8,-2.1) {$W^2$};
  \node at (-2.4,0.8) {(b)};
\end{tikzpicture}
    \caption{Coherent diffractive photoproduction of a quark-antiquark-gluon system in the dipole picture: (a) gluon DTMD term, with a hard quark-antiquark pair and a semi-hard gluon, and (b) quark DTMD term, with a hard quark-gluon pair and a semi-hard antiquark. The zigzag line represents the exchange of a color-neutral pomeron, which generates the rapidity gap $Y_{\mathbb{P}}$. 
}
    \label{fig:diffraction}
\end{figure*}
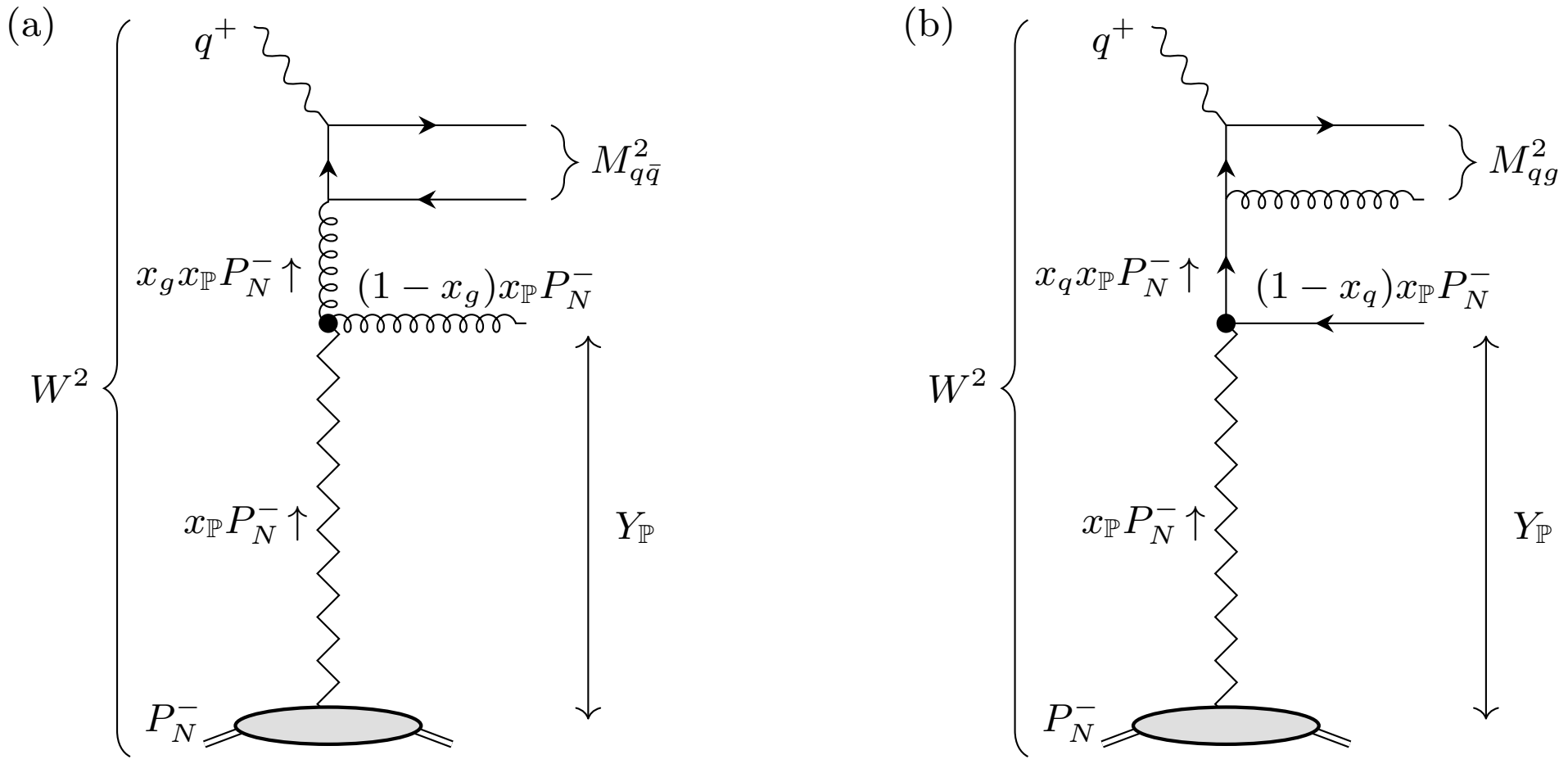

In this section, we describe coherent diffractive charm photoproduction in the CGC framework \cite{Iancu:2003xm}. In the dipole picture, two contributions are relevant: the exclusive process $\gamma+A\to c\bar c+A$ (\fig\ref{fig:exclusive}) and the diffractive process $\gamma+A\to c\bar c g+A$ (\fig\ref{fig:diffraction}). In both, the interaction is mediated by a color-singlet pomeron $(\mathbb{P})$ and the target stays intact after the hard scattering. 

This section is organized as follows. We first define the kinematics common to both processes in Sect.\,\ref{sec:kinematics}, we present the two cross sections in Secs.\,\ref{sec:exclusive} and \ref{sec:diffractiveTMD}, and finally we compare them in Sec.\,\ref{sec:comparison}.

\subsection{Diffractive kinematics}
\label{sec:kinematics}
In light-cone coordinates the real photon and a nucleon of the target nucleus have momenta $q=(q^+,0,\boldsymbol{0})$ and $P_N=(0,\PN,\boldsymbol{0})$, respectively. The produced charm quark has transverse momentum $\ptc=|\pbc|$, rapidity $y$ and plus momentum $p^+=m_Te^{y}/\sqrt2$, where $m_T=\sqrt{m^2+\ptc^2}$ is the transverse mass. We denote by $\zq=p^+/q^+$ its longitudinal momentum fraction relative to the photon. In the nucleon-nucleon 
center-of-mass frame,
\begin{align}
\label{eq:qp*}
    q^+=\frac{p^+}{\zq} =\frac{m_T}{\sqrt{2}\zq}e^y\,, \qquad \PN=\frac{P_N^0-P_N^z}{\sqrt{2}}=\frac{\sqrt{s_{NN}}}{\sqrt{2}}  \,,
\end{align}
with $P_N^0=-P_N^z=\sqrt{s_{NN}}/2$. The photon-nucleon center-of-mass energy squared is
\begin{equation}
\label{eq:W2}
    W^2=(P_N+q)^2=2q^+\PN=\sqrt{2s_{NN}}\,q^+\,.
\end{equation}

The pomeron (zigzag line in Figs.\,\ref{fig:exclusive} and \ref{fig:diffraction}) carries the fraction $\xpo$ of the target minus momentum. Diffractive events are identified by a large rapidity gap $Y_{\mathbb P}=\ln(1/\xpo)$ between the produced system $X$ and the outgoing nucleus, or equivalently, by the invariant mass of $X$,
\begin{equation}
     M_X^2=(\xpo P_N +q)^2\,, \qquad \xpo=\frac{M^2_X}{W^2}\,.
\end{equation}
In the exclusive process, the quark and the antiquark are back-to-back with transverse momenta $\pm\pbc$. In the diffractive process of \fig\ref{fig:diffraction}, a third, semi-hard parton with transverse momentum $\Kt\ll\ptc$ is also produced, and the measured quark forms a hard pair with the antiquark as in \fig\ref{fig:diffraction}-(a) or with the gluon as in \fig\ref{fig:diffraction}-(b). Since the third parton is not measured, $\Kt$ is integrated over, and we neglect it in the kinematics of the hard pair, whose invariant mass is then
\begin{equation}
\label{eq:InvariantMass}
    M^2_{q\bar{q}}=\frac{m_T^2}{\zq(1-\zq)} \,,\qquad M_{qg}^2=\frac{\ptc^2}{\zq(1-\zq)}+\frac{m^2}{\zq}\,.
\end{equation}
The hard pair carries a fraction $x$ of the pomeron minus momentum, denoted by $x_g$ in (a) and $x_q$ in (b). We reconstruct these target momentum fractions from the invariant mass of the hard system in the final state using $M_{q\bar q}^2=(x_g\xpo P_N+q)^2=x_g\xpo W^2$, and similarly for $M_{qg}^2$:
\begin{equation}
\label{eq:relation_x_xpom}
    x_g=\frac{M_{q\bar{q}}^2}{\xpo W^2} \,,\qquad x_q=\frac{M_{qg}^2}{\xpo W^2}\,.
\end{equation}
To get a total diffractive cross section one must integrate over  the pomeron momentum fraction $\xpo$, i.e. the size of the rapidity gap.
Since $x_{g,q}\leq1$, the lower limit for $\xpo$ is given by the kinematics: $\xpo\geq x_{\mathbb{P}\min}$, with $x_{\mathbb{P}\min}=M_{q\bar q}^2/W^2$ for the gluon term and $M_{qg}^2/W^2$ for the quark term. The upper limit depends on the definition of diffractive events chosen in the experimental analysis, which we assume here would be performed by imposing a minimum size for the rapidity gap $Y_{\mathbb P}$. For concreteness, we will here choose $\xpo\leq x_{\mathbb{P}\max}=0.1$, following the choice made in the analysis in Ref.\,\cite{Guzey:2026lfb}.

\subsection{Exclusive quark-antiquark production}
\label{sec:exclusive}
The exclusive cross section in \fig\ref{fig:exclusive} is obtained by including the quark mass in the calculation of Ref.\,\cite{Hauksson:2024bvv}:
\begin{multline}
\label{eq:ExclusiveCrosssection}
    \frac{\der\sigma^\text{exclusive}_{\gamma+A}}{\der y\der^2\pbc}=\frac{\aem e^2_q \Nc S_\perp m^2}{2\pi^2}\\
    \times \zq\Big[\big(\zq^2+(1-\zq)^2\big)\big[\mathcal{Q}_1(\ptc,\xpo)\big]^2+ \big[\mathcal{Q}_0(\ptc,\xpo)\big]^2\Big] \,.
\end{multline}
Here $e_q=2/3$ is the charm-quark charge and $S_\perp$ the transverse area of the target. The antiquark carries the fraction $1-\zq$ of the photon momentum and the transverse momentum $-\pbc$, so the cross section depends only on the quark variables, with $\xpo$ given by \eq\eqref{eq:xpom}. The functions $\mathcal{Q}_{n}$, $n=0,1$, are
\begin{align}
\label{eq:Qn_exclusive}
    \mathcal{Q}_n(\ptc, \xpo)&=\int \der r_\perp\, r_\perp \operatorname{J}_n(\ptc r_\perp)\operatorname{K}_n(mr_\perp)\mathcal{N}(r_\perp,\xpo) \,,
\end{align}
with $\operatorname{J}_n$ and $\operatorname{K}_n$ the Bessel functions of the first kind and modified second kind. In the limit $m\rightarrow 0$, $\operatorname{K}_1(mr_\perp)\rightarrow 1/(mr_\perp)$ and $\operatorname{K}_0(mr_\perp)\rightarrow -\ln(mr_\perp)$, so the second term in \eq\eqref{eq:ExclusiveCrosssection} vanishes and the massless result of Ref.\,\cite{Hauksson:2024bvv} is recovered.

The fundamental representation dipole amplitude $\mathcal{N}(r_\perp,\xpo)$ describes the scattering of a quark-antiquark dipole of size $r_\perp$ off the target,
\begin{align}
\label{eq:DipoleAmplitude}
    \mathcal{N}(r_\perp,\xpo)&=1-D(r_\perp,\xpo) \nonumber \\
    &=1-\frac{1}{\Nc}\big\langle\Tr(V(r_\perp)V^\dagger(0))\big\rangle_{\xpo} \,.
\end{align}
Here $D$ is the dipole scattering matrix where each Wilson line $V$ describes the scattering of the quark or antiquark at a fixed transverse position. The brackets denote the average over the color configurations of the target, taken at the amplitude level for coherent diffraction. Given an initial condition, the dependence of the dipole amplitude on $\xpo$ is obtained by solving the BK equation \cite{Balitsky:1995ub, Kovchegov:1999yj,Balitsky:2006wa}.

In the exclusive process, the quark pair carries all the pomeron minus momentum, $x=1$ and $M_X=M_{q\bar q}$. As a result, the $\xpo$ value is fixed by $q^+$ 
at a given $y$ and $\ptc$ :
\begin{align}
\label{eq:xpom}
    \xpo(q^+)&=\frac{m_T^{2}}{2q^+P_N^-\zq(1-\zq)}=\frac{m_Te^{-y}}{\sqrt{s_{NN}}\left(1-\frac{m_Te^{y}}{\sqrt{2}q^+}\right)}\,.
\end{align}

As discussed in Sec.~\ref{sec:PhotonFlux}, in UPCs the partonic cross sections are integrated over a photon flux $F(\zp)$ 
with 
$q^+=\sqrt2\,\zp E_A$. This integral over $z_\gamma$ must then be performed with a Dirac $\delta$-function enforcing the constraint \eq\eqref{eq:xpom}, which has a unique solution for $z_\gamma$, denoted as $z_\gamma^*$ and given by
\begin{equation}
    \label{eq:zgammastar}
    z_\gamma^*=\frac{\xpo m_Te^{2y}}{\xpo\sqrt{s_{NN}}e^{y}-m_T}\,,
\end{equation}
since $\xpo$ decreases monotonically with the photon energy at a fixed $y$ and $\ptc$. Including the Jacobian of the transformation, one finds
\begin{equation}
\label{eq:exclusiveUPC_lnx}
    \frac{\der\sigma^{\text{exclusive}}_{\text{UPC}}}{\der y\,\der^2\pbc\,\der\ln(\xpo)}
    =\frac{(z_\gamma^*)^2F(z_\gamma^*)}{\xpo e^{2y}} \,\,
    \frac{\der\sigma^{\text{exclusive}}_{\gamma+A}}{\der y\,\der^2\pbc}\,.
\end{equation}

\subsection{Diffractive quark-antiquark-gluon production}
\label{sec:diffractiveTMD}
At leading order in $\as$, the hard pair is accompanied by a semi-hard parton with transverse momentum $\Kt\sim \qs \sim $ GeV. In the correlation limit, where the hard scale $M_h$ is much larger than the saturation scale $\qs$ of the target, the cross section factorizes into hard factors and the DTMDs of the pomeron \cite{Hauksson:2024bvv, Iancu:2022lcw,Caucal:2026xhj}. Two configurations are shown in \fig\ref{fig:diffraction}: (a) a hard quark-antiquark pair with a semi-hard gluon, $M_h=M_{q\bar q}$, and (b) a hard quark-gluon pair with a semi-hard antiquark, $M_h=M_{qg}$. At $\zq=1/2$, these are $M_{q\bar q}= 2m_T$ and  $M_{qg}\approx2\ptc$. 

At fixed photon energy, the differential cross section for a quark produced with transverse momentum $\pbc$ and rapidity $y$ is the sum of the two contributions \cite{Caucal:2026xhj}:
\begin{multline}
\label{eq:DiffractiveCS_lnx}
    \frac{\der\sigma^{\text{diffractive}_{\text{TMD}}}_{\gamma+A}}{\der y\,\der^2\pbc\,\der\ln(\xpo)}
    =\mathcal{H}(\zq,\ptc)\int\der^2\mathbf{K}\;g_\po(x_g,\Kt,\xpo)\\
    +\mathcal{H}'(\zq,\ptc)\int\der^2\mathbf{K}\;q_{\mathbb{P}}(x_q,\Kt,\xpo) \,.
\end{multline}
Although the transverse momentum imbalance $\Kt$ of the hard pair, which is determined by the transverse momentum of the additional semi-hard parton, is integrated out, we refer to this cross section as ``diffractive$\tmd$'' to emphasize its origin in a TMD-factorized cross section. The first term (a) contains the gluon DTMD $g_\po$ and the second term (b) the quark DTMD $q_\po$. 
In UPCs, \eq\eqref{eq:DiffractiveCS_lnx} is integrated over $\zp$ including the flux $F(\zp)$. In future measurements, the cross section will likely be integrated over a finite range of $\xpo$. Accordingly, \eq\eqref{eq:DiffractiveCS_lnx} should also be integrated over $\xpo$.

\paragraph*{(a) Gluon term.}
Here the quark and the antiquark are not back-to-back: their momentum imbalance $\mathbf{K}$ is balanced by the gluon, which carries a small longitudinal momentum fraction, so the antiquark still carries $1-\zq$. The hard factor describes the splitting of a real, transversely polarized photon into a quark-antiquark pair that absorbs a gluon from the pomeron,
\begin{multline}
\label{eq:hard factor}
    \mathcal{H}(\zq,\ptc)\equiv
   \aem \as e_q^2\zq\Bigg[\big( \zq^2+ (1-\zq)^2  \big) \frac{\ptc^4 + m^4}{\big(\ptc^2 + m^2\big)^4}\\
   + 2m^2  \frac{\ptc^2}{ \big( \ptc^2+m^2\big)^4}\Bigg] \,.
\end{multline}
Here $\as$ is the one-loop running coupling which we evaluate at a scale $\mu_R$ set by the charm transverse mass $\mu_R=m_T$. The gluon DTMD is the number density of gluons in the pomeron with transverse momentum $\Kt$ and momentum fraction $x$,
\begin{equation}
\label{eq:GDTMD}
    g_\po (x,\Kt,\xpo)\equiv \frac{S_\perp(\Nc^2-1)}{8\pi^4(1-x)}[G(x,\Kt,\xpo)]^2 \,,
\end{equation}
where
\begin{multline}
\label{eq:G}
    G(x,\Kt,\xpo)=\Omega^2\int_0^\infty\der R_\perp\, R_\perp J_2(\Kt R_\perp)
    K_2(\Omega R_\perp)\\
    \times\mathcal{N}_g(R_\perp,\xpo)\,.
\end{multline}
Here $\Omega^2=x\Kt^2/(1-x)$ and $\mathcal{N}_g=2\mathcal{N}-\mathcal{N}^2$ is the adjoint representation gluon-gluon dipole amplitude in the large-$\Nc$ limit, with $\mathcal{N}$ the quark dipole amplitude in \eq\eqref{eq:DipoleAmplitude}.

\paragraph*{(b) Quark term.}
In the second term, the measured quark forms the hard pair with the gluon, and the antiquark is semi-hard\footnote{The case of a measured antiquark with a semi-hard quark gives the same contribution by charge conjugation.} \cite{Caucal:2026xhj,Hauksson:2024bvv}. For a real photon, the hard factor is \cite{Caucal:2026xhj}
\begin{equation}
\label{eq:hard factor quark}
    \mathcal{H}'(\zq,\ptc)\equiv 2\,\aem\as e_q^2\,C_F\,\zq\;\frac{\zq\,\big(1+(1-\zq)^2\big)}{\ptc^4}\,,
\end{equation}
where $C_F=(\Nc^2-1)/(2\Nc)$ is the fundamental quadratic Casimir factor. The quark mass is here a semi-hard scale and enters only the quark DTMD,
\begin{multline}
\label{eq:qDTMD}
    q_\po(x,\Kt,\xpo)\equiv \frac{S_\perp\Nc}{8\pi^4}\,\frac{x}{1-x}\,\frac{\Kt^2+m^2}{x\Kt^2+m^2}\\
    \times\left[\tilde{\mathcal{Q}}_1^2(x,\Kt,\xpo)+\frac{(1-x)\,m^2}{x\Kt^2+m^2}\,\tilde{\mathcal{Q}}_0^2(x,\Kt,\xpo)\right],
\end{multline}
where
\begin{multline}
\label{eq:Qn}
    \tilde{\mathcal{Q}}_n(x,\Kt,\xpo)=\tilde\Omega^2\int_0^\infty\der R_\perp\,R_\perp\,J_n(\Kt R_\perp)\\
    \times K_n(\tilde\Omega R_\perp)\,\mathcal{N}(R_\perp,\xpo)\,,
\end{multline}
for $n=0,1$. Here $\tilde\Omega^2=(m^2+x\Kt^2)/(1-x)$ reduces to $\Omega^2$ for $m=0$. Unlike $G$ in \eq\eqref{eq:G},  $\tilde{\mathcal{Q}}_n$ contains the quark dipole amplitude: the semi-hard antiquark and the compact quark-gluon pair scatter as a fundamental representation quark-antiquark dipole.

The TMD-factorized cross section \eqref{eq:DiffractiveCS_lnx} is valid for $\Kt\ll M_h$. We therefore integrate the DTMDs from $\Kt=0$ up to the hard scale of each term~\cite{Caucal:2026xhj}. For the gluon term, the upper limit is $M_{q\bar q}$ at the given value of $\zq$, which is at least $2m_T$, the value reached at $\zq=1/2$. Since the gluon DTMD is approximately constant for $\Kt\lesssim\qs$ and falls like $\qs^4/\Kt^4$ above it~\cite{Iancu:2022lcw,Hauksson:2024bvv}, the integral is dominated by $\Kt\sim\qs$, and its dependence on the upper limit is of relative order $\qs^2/M_h^2$. This correction is small at large $\ptc$, but not for $\ptc\lesssim2$~GeV, where $2m_T$ is roughly only 2 times the saturation scale of the gluon dipole, so that our diffractive$\tmd$ results in this region are only marginally within the validity of the DTMD factorization. For the quark term, we integrate up to $2\ptc$, beyond which the result no longer depends on the cut, and we include this term only for $\ptc>5$~GeV, where $M_{qg}\approx2\ptc$ is large compared to both $\qs$ and the charm mass. As we show in Sec.~\ref{sec:comparison}, it contributes only close to $x_{\mathbb{P}\min}$, and its contribution to the diffractive$\tmd$ cross section integrated over $\xpo$ is much smaller than that of the gluon term.

In our analysis, the $x$, $\xpo$ and $m_T$ dependence of the $\Kt$-integrated~\mbox{DTMDs}, i.e.~the diffractive PDFs (DPDFs), is obtained directly from their CGC expressions \eqs\eqref{eq:GDTMD}-\eqref{eq:qDTMD}. In this approach, the $m_T$ (or $p_\perp$) dependence enters only through the upper boundary of the $\Kt$ integration (as just discussed), while the $\xpo$ dependence arises from the BK evolution of the dipole amplitude. This treatment, however, neglects the modifications induced by DGLAP evolution in the $x$ and hard-scale dependence of the DPDFs. This effect has been studied in~\cite{Iancu:2022lcw,Iancu:2023lel,Hauksson:2024bvv,Iancu:2025jsu} for the gluon and quark DPDFs and should be incorporated in a future refinement of our analysis to provide a more complete description of the DPDFs at large $M_h$. Nevertheless, we expect this omission to have a limited impact on our results, since the dominant source of $p_\perp$ dependence comes from the hard factors in \eq\eqref{eq:DiffractiveCS_lnx}. Including DGLAP evolution would introduce an additional, relatively mild logarithmic dependence on $p_\perp$ in the DPDFs, inherited from the DGLAP evolution of the CGC expressions used here.

\subsection{Exclusive versus diffractive\texorpdfstring{$_\text{TMD}$}{-TMD} production}
\label{sec:comparison}
\begin{figure}
    \centering
    \includegraphics[width=1\linewidth]{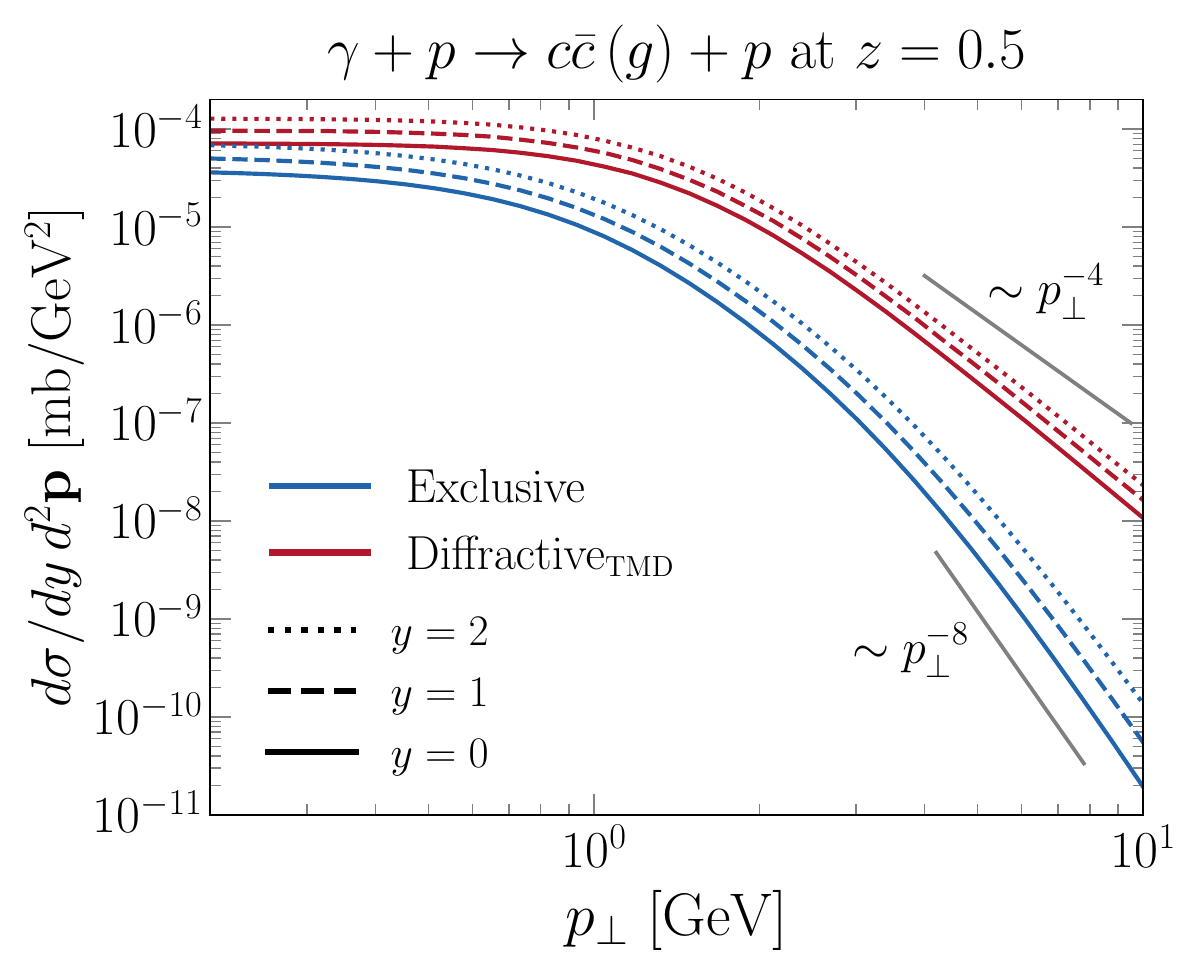}
    \caption{Charm photoproduction off a proton at $\sqrt{s_{NN}}=5.36$~TeV in the exclusive (blue) and diffractive$_\text{TMD}$ (red) channels as a function of the charm quark transverse momentum $\ptc$ at several values of the rapidity $y$. The photon energy is fixed at $q^+ = 2p^+$ ($\leftrightarrow$ $z = 0.5$). By gray lines we indicate the power-law behavior of the two channels at large $\ptc$.}
    \label{fig:partonCS}
\end{figure}
In the CGC power counting, what we refer to in this paper as the exclusive process is the leading-order contribution in $\as$. A full NLO calculation (see~\cite{Fucilla:2023mkl} for massless quarks) would include gluon emission in exact kinematics, while the diffractive TMD limit that we are using would be classified as a partial NLO result which is expected to dominate in the hard scattering kinematics. Before moving to UPC cross sections, let us here quantify these contributions at the parton level.

In \fig\ref{fig:partonCS} we compare the exclusive and diffractive contributions at fixed photon energy, using the BK-evolved dipole amplitude   discussed in more detail in Sec.~\ref{sec:dipole}. We observe that in the kinematics studied in this  manuscript, the exclusive cross section falls approximately as $\ptc^{-8}$, while the diffractive$\tmd$ scales as $\ptc^{-4}$, because of the large $p_\perp$ behavior of the hard factor $\mathcal{H}$ and $\mathcal{H}'$ in \eqs\eqref{eq:hard factor}-\eqref{eq:hard factor quark}. Therefore, the latter dominates at large $\ptc$ although it is formally suppressed by $\as$. The physical explanation is that, in the exclusive process, the target must transfer the full momentum $\ptc$ to a small dipole of size $r_\perp\sim1/\ptc$, which scatters weakly due to color transparency. In contrast, in the diffractive$\tmd$ process the recoil is taken by the gluon or antiquark at a large distance $R_\perp\sim1/\qs$, where the scattering is strong. In the approach we use here, the diffractive contribution is also the largest at low $\ptc$. However, in this regime virtual contributions that are part of the full NLO result and are not included in the TMD limit used  can be expected to be large. Thus a more reliable calculation at small $\ptc$ would require a full NLO calculation, which is outside the scope of this study. So far such a calculation has only been performed for massless quarks, see Refs.~\cite{Beuf:2024msh,Kaushik:2025roa}.

Let us now discuss the high-$\ptc$ asymptotics of the exclusive cross section in more detail. 
At large $\ptc$, the integrals $\mathcal{Q}_{0,1}$ in \eq\eqref{eq:Qn_exclusive} are dominated by small dipoles, $r_\perp\sim1/\ptc$. In the MV model, which is the initial condition for our BK-evolved amplitudes, the small dipole behavior is $\mathcal{N}\propto r_\perp^2\qs^2\ln(1/(r_\perp\Lambda_{\text{QCD}}))$. The logarithmic dependence on $r_\perp$ gives rise to a power-law tail in the Bessel-function integrals. The dominant asymptotic behavior $\sim \qs^4/\ptc^6$, resulting from the $\mathcal{Q}_1$ term, is the same as for massless quarks in DIS~\cite{Iancu:2022lcw,Hauksson:2024bvv}.\footnote{The logarithm in the MV model, or anomalous dimension in a BK-evolved dipole amplitude, is essential for understanding the large-$\ptc$ behavior. If, as in the Golec-Biernat--Wusthoff parametrization~\cite{Golec-Biernat:1998zce}, the dipole amplitude behaves precisely as $\mathcal{N}\propto r_\perp^2\qs^2$ at small $r_\perp$, there is a cancellation of the leading terms in $\mathcal{Q}_1$, resulting in $\mathcal{Q}_0$ dominating the large-$\ptc$ behavior. As a consequence the high-$\ptc$ asymptotics becomes proportional to the quark mass $m^2\qs^4/\ptc^8$. Such a behavior is, however, quite fortuitous and generically one would expect the large $\ptc$ behavior to be the same as for massless quarks.}
Numerically, however, this asymptotic behavior  is only reached for a fixed, unevolved, amplitude at higher $\ptc$ than what we are plotting.

In practice, for values of $\ptc$ that are relevant for LHC phenomenology,  our spectrum remains closer to $\ptc^{-8}$. This is due to several effects. First, for the HERA fit that we are using (our initial condition is the so called MV$^e$ parametrization,  see Eq.~\eqref{eq:ICBK_proton} with $e_c=18.9$), the logarithm becomes important only for small dipoles, $r_\perp\lesssim1/(e_c e\Lambda_{\text{QCD}})$, that is, for $\ptc\gtrsim12$~GeV. Secondly,  going to higher $\ptc$ for fixed $y$, one goes to earlier evolution rapidities, where $\qs$ gets smaller. This  makes the spectrum in $\ptc$ steeper.  In summary, the $\ptc^{-8}$ behavior observed in  \fig\ref{fig:partonCS} is not a generic fixed power, but a combination of different effects and specific to this kinematic regime and   dipole amplitude parametrization. The essential feature remains that the exclusive contribution is suppressed compared to the diffractive one.

In \fig\ref{fig:partonCS}, both cross sections increase with $y$ at fixed $z=1/2$. For the exclusive process, increasing $y$ at fixed $z=1/2$ and $p_\perp$ leads to a decrease in $\xpo=2m_T e^{-y}/\sqrt{s_{NN}}$, according to \eq\eqref{eq:xpom}. This results in a larger dipole amplitude due to small-$\xpo$ evolution, explaining the $y$ dependence of the blue curves in \fig\ref{fig:partonCS}. The situation is more involved for the diffractive TMD contribution, since the rapidity $y$ enters through the momentum fractions $x_g$ and $x_q$ of the gluon and quark DTMDs, respectively, while $\xpo$ is integrated over. As we shall see shortly, the diffractive TMD cross section is dominated by the gluon DTMD contribution. As noted in~\cite{Iancu:2022lcw}, the gluon DTMD $g_{\mathbb{P}}$ divided by $1-x$ approximately exhibits geometric scaling as a function of $K_\perp^2/\tilde Q_s^2(\xpo,x)$, with $\tilde Q_s^2(\xpo,x)=(1-x)Q_s^2(\xpo)$, even when BK $\xpo$-evolution is included. Since the $\Kt$ integral of $g_{\mathbb{P}}$ is dominated by the saturation regime, it therefore scales approximately as $(1-x)\tilde Q_s^2$, which increases as $x$ decreases. At fixed $\xpo$, increasing $y$ indeed leads to a decrease in $x_g$, as follows from $x_g=2m_T e^{-y}/(\xpo\sqrt{s_{NN}})$ (for $z=1/2$). Consequently, the diffractive TMD contribution also increases with $y$ after integration over $\xpo$, as illustrated by the red curves in \fig\ref{fig:partonCS}.

\begin{figure}
    \centering
    \includegraphics[width=1\linewidth]{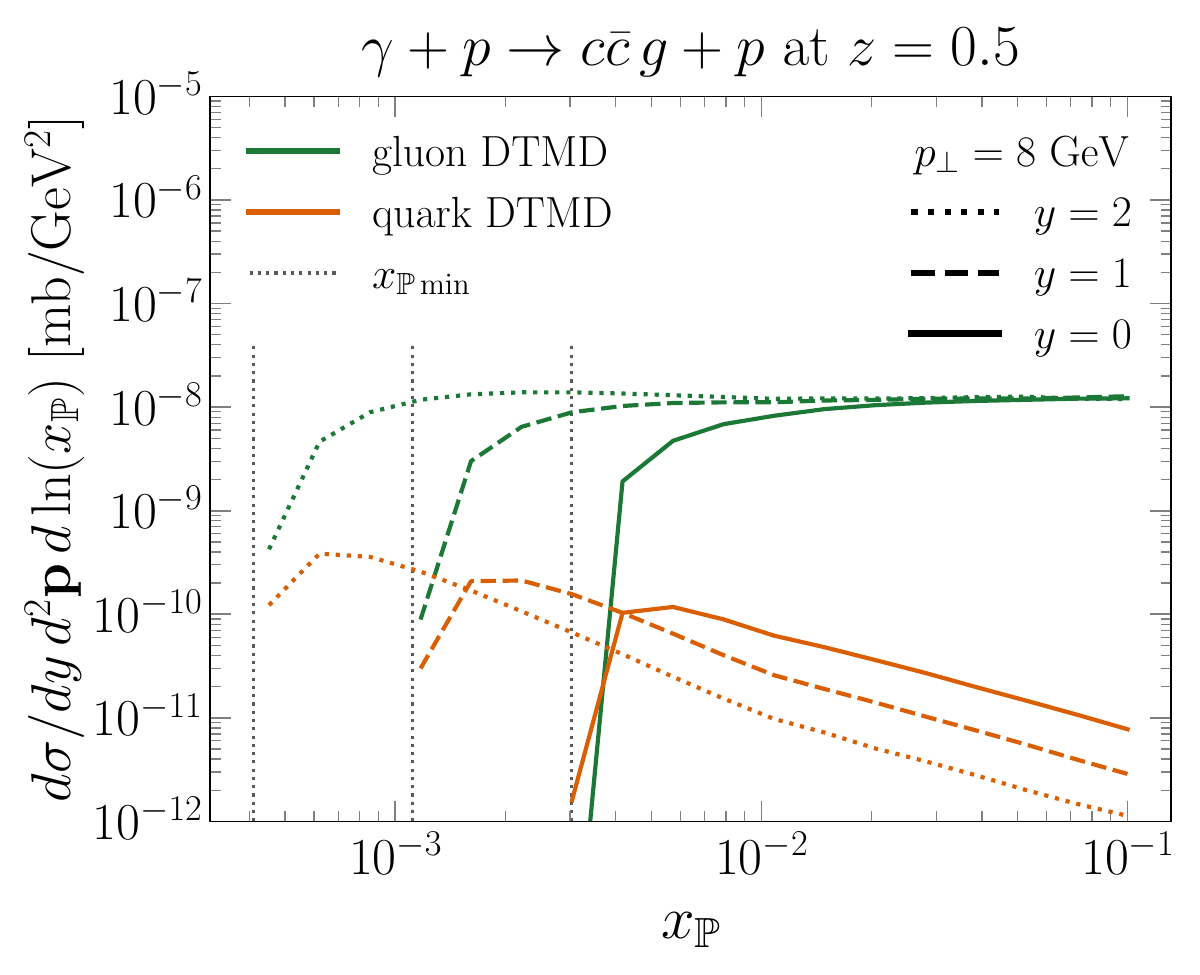}
    \caption{Gluon (green) and quark (orange) DTMD terms of the diffractive$_\text{TMD}$ cross section as a function of $\xpo$, for the rapidities $y=0$ (solid), $1$ (dashed) and $2$ (dotted). We fix the photon energy $q^+ = 2p^+$ (equivalently $z = 0.5$) and the quark transverse momentum $\ptc = 8$~GeV at $\sqrt{s_{NN}} = 5.36$ TeV, for a proton target. The vertical dotted lines show $x_{\mathbb P\min} = M_{q\bar q}^2/W^2$ for each rapidity. 
    }
    \label{fig:xpom}
\end{figure}

For the diffractive TMD cross sections, it is also instructive to show the cross section differentially in $\xpo$, at fixed $z$ and $p_\perp$, for several values of $y$. This is done in \fig\ref{fig:xpom}, where we show the distribution in $\xpo$ for several rapidities at a fixed $\ptc=8$~GeV. We also show the minimal value 
$x_{\mathbb{P}\min}=2m_Te^{-y}/\sqrt{s_{NN}}$, obtained by setting $z=1/2$ in \eq\eqref{eq:xpom}. The green curves, corresponding to the gluon contribution, are in qualitative agreement with the $(1-x_g)\tilde Q_s^2(\xpo,x_g)$ scaling discussed in the previous paragraph: (i) at fixed $\xpo$, the cross-section increases as $y$ increases, (ii) at fixed $y$, the cross section starts at  $x_{\mathbb{P}\min}$ and is nearly flat above it, since for $\xpo\gg x_{\mathbb{P}\min}$ one has $x\ll1$, where the gluon DTMD is independent of $x$~\cite{Iancu:2022lcw} as a consequence of the scaling with $\tilde Q_s^2(\xpo,x)\simeq Q_s^2(\xpo)$. Above $x_0=0.01$, the dipole amplitude is frozen and the gluon DTMD is constant. Below $x_0=0.01$, it grows mildly due to the BK evolution which increases the value of $Q_s^2(\xpo)$. The integrated cross section is therefore roughly proportional to the size of the $\xpo$ range, which grows with $y$ and depends on the cut that we have chosen as $x_{\mathbb{P}\max}=0.1$.

On the other hand, the quark DTMD shown in the orange curves in \fig\ref{fig:xpom} is proportional to $x\propto1/\xpo$ at small $x$ \cite{Caucal:2026xhj}. The quark term therefore contributes only in a narrow region just above $x_{\mathbb{P}\min}$. This is mainly because of the charm mass: since $\tilde\Omega\geq m$ in \eq\eqref{eq:Qn}, the semi-hard antiquark is restricted to small dipoles, $R_\perp\lesssim1/m$, which scatter weakly. This leads to an overall $(Q_s/m)^4$ suppression of the charm quark occupation number in the nucleus~\cite{Caucal:2026xhj}. The massless gluon dipole, instead, probes $R_\perp\sim1/\qs$. In addition, the quark dipole amplitude $\mathcal{N}$ is smaller than $\mathcal{N}_g$ because the saturation scale squared in the gluon-gluon dipole amplitude is larger than in the quark-antiquark dipole amplitude by a factor $C_A/C_F$, which becomes $2$ in the large-$\Nc$ form $\mathcal{N}_g=2\mathcal{N}-\mathcal{N}^2$ used here.  As a result, the quark term is less than $1\%$ of the diffractive$_\text{TMD}$ cross section. 

\begin{figure*}
    \centering
    \includegraphics[width=0.8\linewidth]{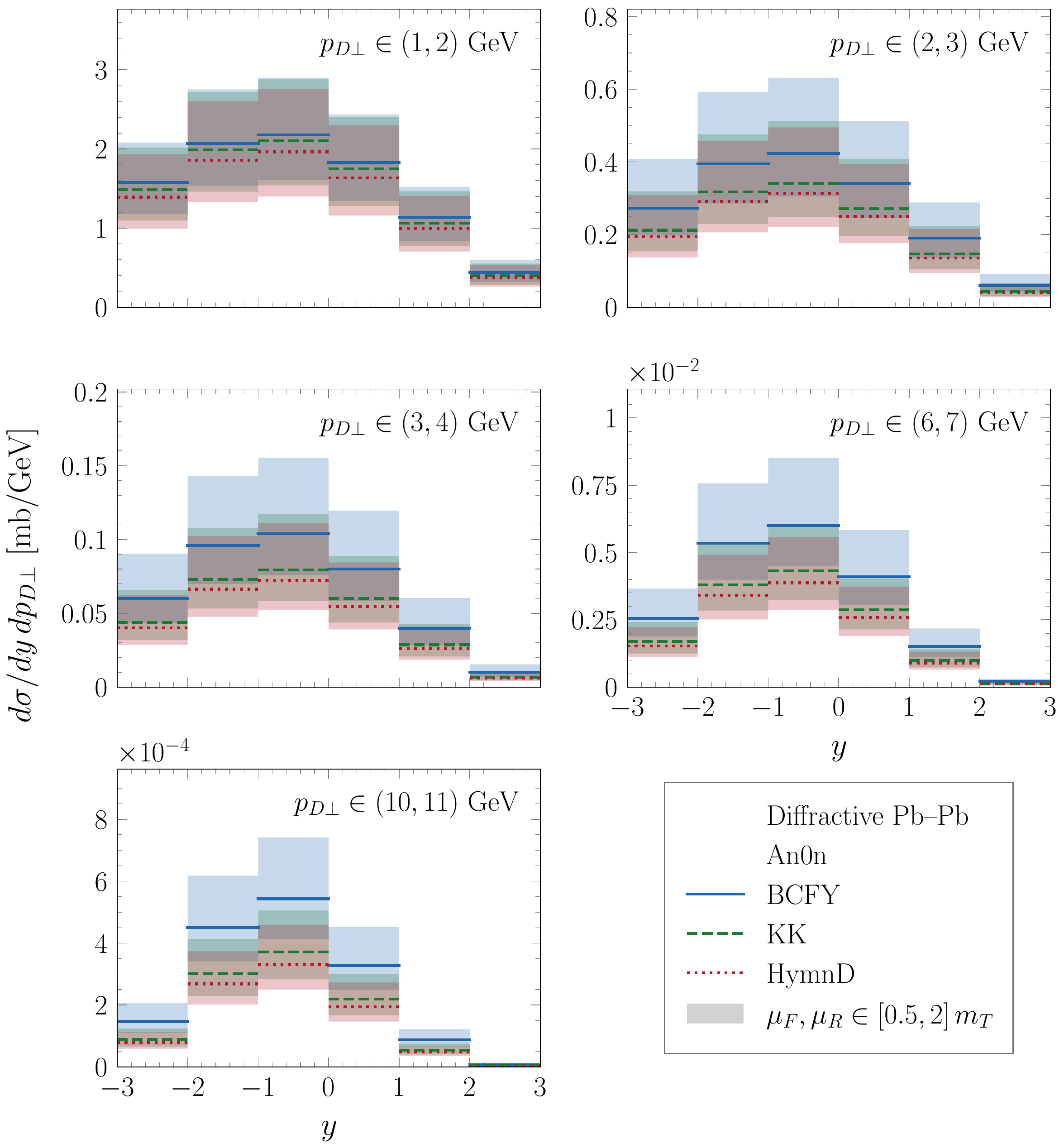}
    \caption{{Coherent diffractive $D^0$ photoproduction cross section \eqref{eq:diffractiveCS} in Pb--Pb UPCs at $\sqrt{s_{NN}}=5.36$ TeV in the $An0n$ neutron class, as a function of the $D^0$ rapidity $y$ in bins of $p_{D\perp}$. The lines show the BCFY (solid), Kniehl--Kramer (dashed) and HymnD (dotted) FF, with the bands given by the seven-point variation of the factorization ($\mu_F$) and renormalization ($\mu_R$) scales, $\mu_F,\mu_R\in[0.5,2]\,m_T$ with $1/2\le\mu_F/\mu_R\le 2$. Positive rapidity is the direction of the photon.}
\label{fig:D0_An0n_PbPb}}
\end{figure*}
\begin{figure*}
    \centering
    \includegraphics[width=0.8\linewidth]{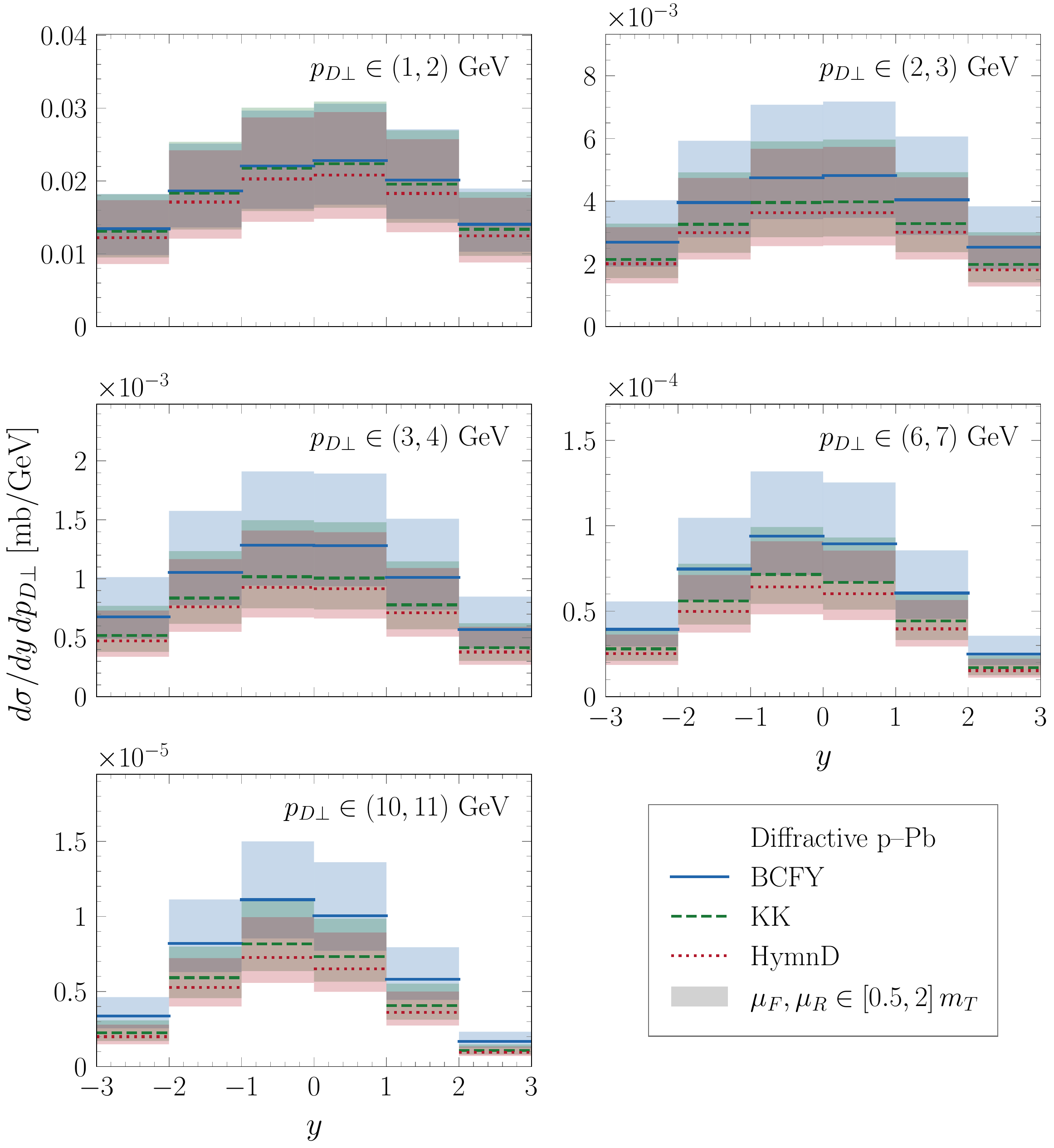}
    \caption{Coherent diffractive $D^0$ photoproduction cross section \eqref{eq:diffractiveCS} in p--Pb UPCs at $\sqrt{s_{NN}}=8.16$ TeV as a function of the $D^0$ rapidity $y$ in bins of $p_{D\perp}$. The lines show the BCFY (solid), Kniehl--Kramer (dashed) and HymnD (dotted) FF, with the bands given by the same scale variation as in \fig\ref{fig:D0_An0n_PbPb}. Positive rapidity is the direction of the photon emitted by the lead.}
    \label{fig:D0_bins_pPb}
\end{figure*}

\section{\texorpdfstring{$D^0$}{D0} photoproduction in UPC\texorpdfstring{\NoCaseChange{s}}{s}}
\label{sec:numerics}

\subsection{Diffractive \texorpdfstring{$D^0$}{D0} in Pb--Pb and p--Pb UPCs}
The differential cross section for diffractive $D^0$ photoproduction in UPCs is a convolution of
the effective photon flux \eq\eqref{eq:eff_photon_flux},   the parton cross sections in \eqs\eqref{eq:ExclusiveCrosssection} and
\eqref{eq:DiffractiveCS_lnx} and the fragmentation function $D(z_h)$:
\begin{multline}
\label{eq:diffractiveCS}
    \frac{\der\sigma_{\text{UPC}}^{\text{diffractive}}}{\der y\,\der^2\mathbf{p}_{D}}
    =\int_0^1\frac{\der z_h}{z_h^2}\,D(z_h)
    \int_{z_\gamma^{\min}}^{1}\der z_\gamma\,F(z_\gamma)\\
    \times
    \Bigg[\int_{x_{\mathbb{P}\min}(z_\gamma)}^{x_{\mathbb{P}\max}}\frac{\der\xpo}{\xpo}\,
    \frac{\der\sigma_{\gamma+A}^{\text{diffractive}_{\text{TMD}}}}{\der y\,\der^2\pbc\,\der\ln\xpo}\\
    +\Theta\big(x_{\mathbb{P}\max}-\xpo(z_\gamma)\big)\,
    \frac{\der\sigma_{\gamma+A}^{\text{exclusive}}}{\der y\,\der^2\pbc}\Bigg]\,.
\end{multline}
Here $\pbc=\mathbf{p}_D/z_h$ is the transverse momentum of the charm quark, which gives the Jacobian $1/z_h^2$. The photon flux $F(z_\gamma)$ is defined in \eq\eqref{eq:eff_photon_flux} for nucleus-nucleus collisions and in \eq\eqref{eq:pAphotonFlux} for proton-nucleus collisions, and the minimum energy $z_\gamma^{\min}$ needed to produce a charm quark is given in \eq\eqref{eq:qmin}. In the
TMD contribution, $x_{\mathbb{P}\min}(z_\gamma)$ is the kinematic threshold of Sec.~\ref{sec:kinematics}, that is, $M_{q\bar q}^2/W^2$ for the gluon term and $M_{qg}^2/W^2$ for the quark term, where $W^2$ is given by $z_\gamma$ through \eq\eqref{eq:W2}. In the
exclusive contribution,  instead, there is no $\xpo$ integral, since $\xpo(z_\gamma)=M_{q\bar q}^2/W^2$ is fixed by the photon energy through \eq\eqref{eq:xpom}. Therefore, the step function $\Theta$ only imposes $\xpo\leq x_{\mathbb{P}\max}$. 
If needed, the integration over $z_\gamma$ for the exclusive contribution can be rewritten as an integration over $\xpo$, using the distribution differential in $\xpo$ given in \eq\eqref{eq:exclusiveUPC_lnx}.

The uncertainty bands in the following figures show the variation of the factorization $(\mu_F)$ and renormalization $(\mu_R)$ scales, as described in Sec.~\ref{sec:FF}. Since the fragmentation functions are
defined only above $\mu_0$, the factorization scale is frozen at $\mu<\mu_0$. On the other hand, the uncertainty from the dipole amplitude is small, $1$--$4\%$ for the exclusive and below $0.5\%$ for the diffractive$\tmd$ contribution, and therefore it is only included in the
nuclear modification ratio. The diffractive$\tmd$ contribution is, however, sensitive to the choice of $x_{\mathbb{P}\max}$, because its distribution in $\ln\xpo$ is nearly flat up to this limit (see Sec.~\ref{sec:comparison}). For example, at $p_{D\perp}=2$~GeV, lowering $x_{\mathbb{P}\max}$ to $0.05$ reduces it by about $10\%$ at $y=3$, $25\%$ at $y=0$ and $40\%$ at $y=-2$, whereas the exclusive contribution changes by less than $7\%$. As discussed above, the appropriate value depends on cuts used in a future experimental analysis. In the following, we use $x_{\mathbb{P}\rm max}=0.1$  as in Ref.~\cite{Guzey:2026lfb}.

In \fig\ref{fig:D0_An0n_PbPb} we show the prediction for diffractive $D^0$ photoproduction \eq\eqref{eq:diffractiveCS} in Pb--Pb UPCs in the $An0n$ class, in bins of the meson transverse momentum $p_{D\perp}$ and rapidity $y$. The cross section is largest for $-2<y<0$ and decreases towards positive rapidity, because more energetic photons are needed there, and thus the cross section is suppressed by the photon flux. Note that we are here extending the calculation to backward rapidities, where the cross section is dominated by values of $\xpo$ close to the upper limit $\xpo \leq 0.1$ that we chose, especially for higher $p_{D\perp}$. This is a region where the dipole amplitude is frozen at its initial condition, since $\xpo > x_0 = 0.01$. The cross section at positive $y$ is more properly in the regime of validity of the small-$x$ framework used here.

Regarding the fragmentation functions, in the lowest bin, $p_{D\perp}\in[1,2]$ GeV, the KK FF, which has a larger normalization, gives a cross section similar to BCFY, and the three predictions agree within $15\%$. At higher $p_{D\perp}$, however, BCFY gives the largest cross section, since the $D^0$ mesons then come mostly from large $z_h$.

To test the sensitivity to the photon flux, we have also repeated the calculation with the STARlight EMD probability in the $An0n$ class of Refs.\,\cite{Eskola:2024fhf,Klein:2016yzr}, instead of the EMD parametrization in \eq\eqref{eq:EMD}. In this case, the results change by less than $1\%$ in every bin, see Appendix~\ref{sec:flux_approximations} for a discussion.

These results can be compared with a recent paper~\cite{Guzey:2026lfb}, where the same bins were computed in the G$\gamma$A--FONLL framework with the same charm mass, the BCFY FF and the cut $\xpo<0.1$. Overall, our results are closer to the ``high shadowing'' scenario of Ref.~\cite{Guzey:2026lfb}. This is consistent with the observation that here the relevant configurations in the photon are large adjoint dipoles. While they are larger than just a quark-antiquark dipole in the exclusive process, they still interact with perturbative semihard cross sections, similarly to the relevant states in the ``high shadowing'' scenario. The ``low shadowing'' scenario, on the other hand, features a stronger nuclear suppression of diffraction driven by meson-like configurations that interact with larger nonperturbative hadronic cross sections~\cite{Frankfurt:2011cs}.
Our $p_{D\perp}$ spectrum is, however, harder. With BCFY, our result for $p_{D\perp}\in[1,2]$~GeV is close to the central value of the ``high shadowing'' scenario. For $2<p_{D\perp}<4$~GeV, it is close to the lower edge of the ``high shadowing'' band. At 6--7~GeV, we are inside this band for $y<0$ and close to its lower edge for $y>0$. At 10--11~GeV, we are up to about 60\% above it for $y<0$ and at its upper edge for $y>0$. In contrast, the rapidity dependence is similar in the two calculations.

Figure~\ref{fig:D0_bins_pPb} shows the corresponding predictions for p--Pb UPCs at $\sqrt{s_{NN}}=8.16$~TeV, where the photon emitted by the lead nucleus moves towards positive rapidity. The rapidities are given in the nucleon--nucleon center-of-mass frame, which is shifted by $|\Delta y|=\frac12\ln(208/82)\approx0.47$ relative to the laboratory frame.
Compared with Pb--Pb collisions, the cross section is about two orders of magnitude smaller at negative and central rapidity, but it falls more slowly towards positive rapidity. Note that in p--Pb collisions the photon flux of the lead nucleus does not involve the EMD probability. Apart from this, the dependence on the fragmentation function and the scale uncertainty are similar to those in Pb--Pb. 

For p--Pb, the predictions of Ref.\,\cite{Guzey:2026lfb} use proton diffractive PDFs fitted to HERA data. 
The two calculations agree within scale uncertainties in most bins. However, our $p_{D\perp}$ spectrum is again slightly harder: for $-1<y<0$, our central value moves from the lower edge of their band at $2$--$4$~GeV to its upper edge at $10$--$11$~GeV.

\begin{figure}
    \centering
    \includegraphics[width=1\linewidth]{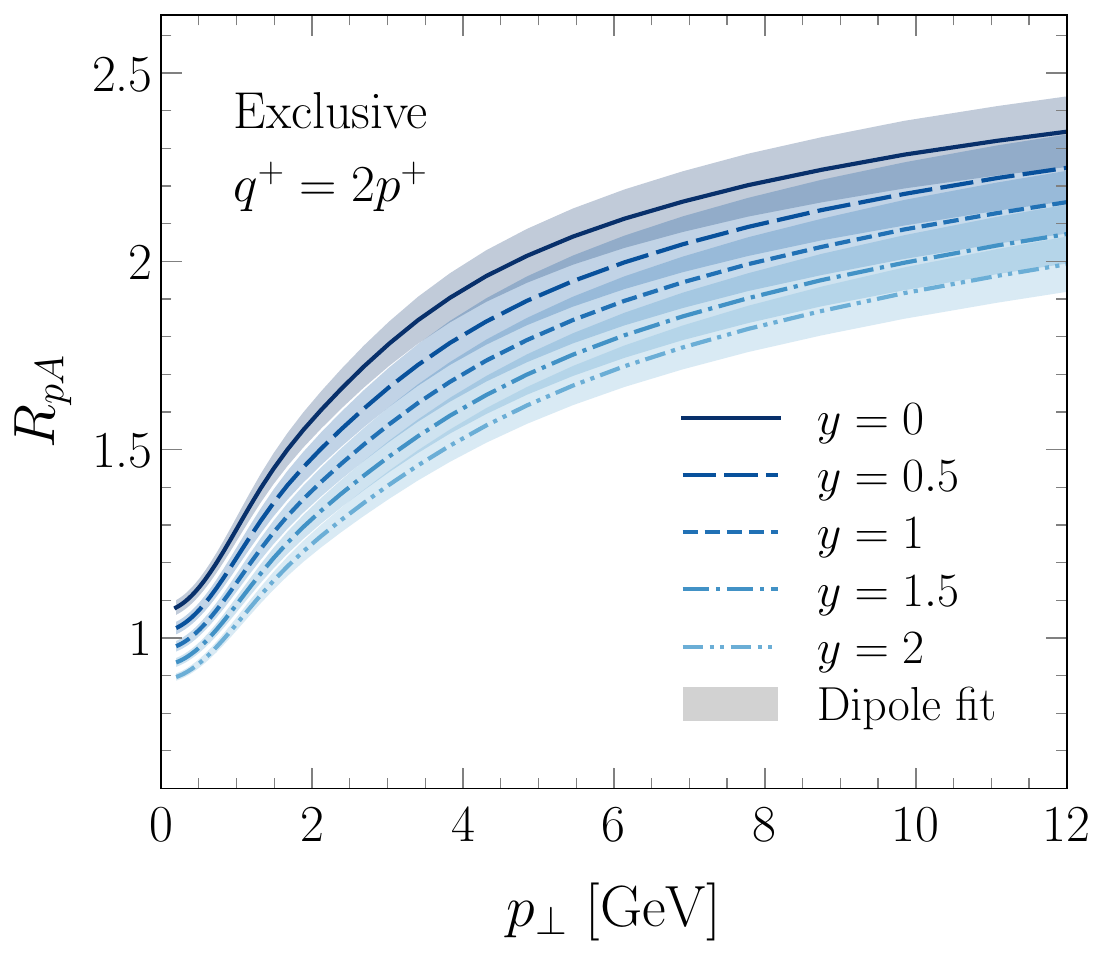}
    \caption{Nuclear modification factor $R_{pA}$ of the exclusive contribution \eq\eqref{eq:ExclusiveCrosssection} for a charm quark with transverse momentum $\ptc$, at fixed photon energy $q^+=2p^+$ and several rapidities. The uncertainty band is the relative spread over the posterior samples of the BK initial condition fitted to HERA data \cite{Casuga:2023dcf}.}
    \label{fig:RpA_exclusive}
\end{figure}
\begin{figure}
    \centering
    \includegraphics[width=1\linewidth]{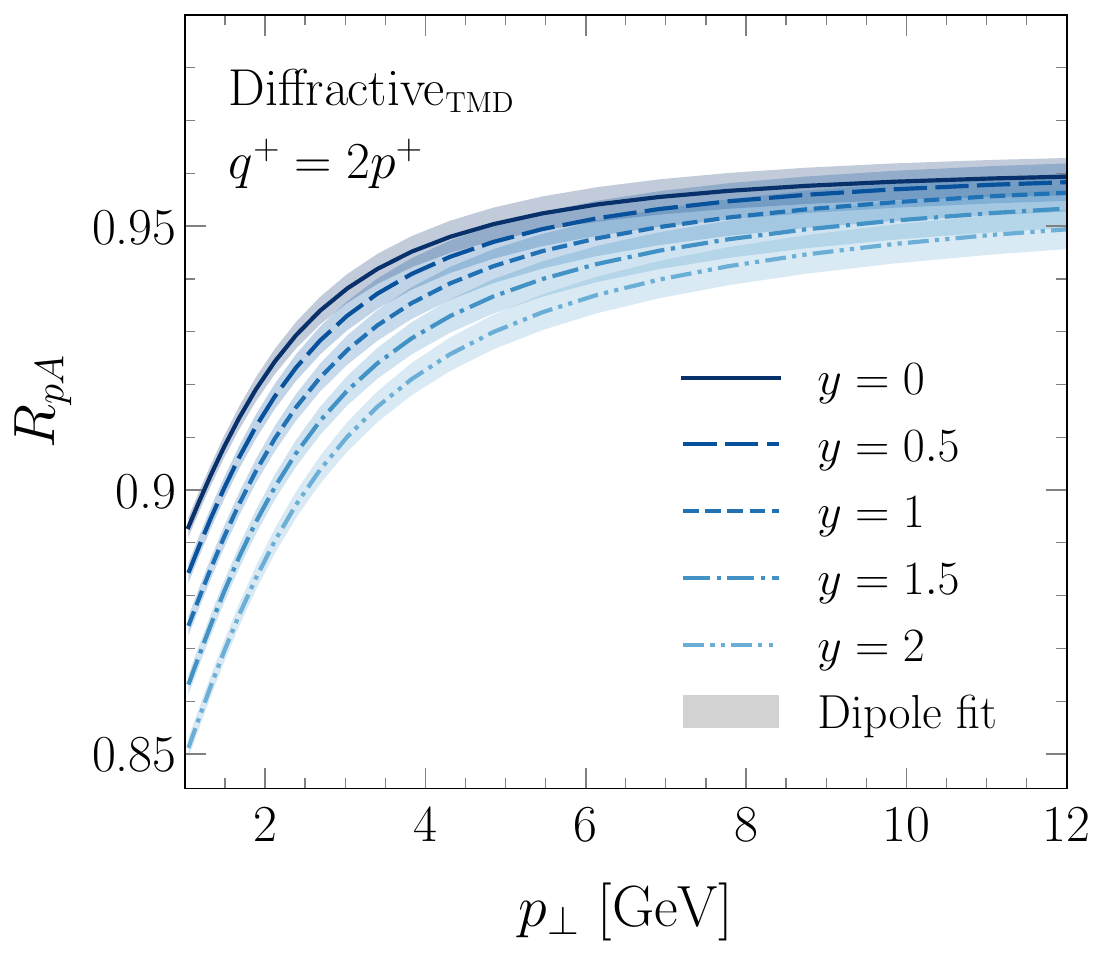}
    \caption{Nuclear modification factor $R_{pA}$ of the diffractive$\tmd$
     contribution at fixed photon energy $q^+=2p^+$ and several rapidities, for $\ptc\geq1$~GeV. The uncertainty band is the relative spread over the posterior samples of the BK initial condition fitted to HERA data \cite{Casuga:2023dcf}.}
    \label{fig:RpA_diffractive}
\end{figure}
To study the nuclear effects, in Figs.~\ref{fig:RpA_exclusive} and \ref{fig:RpA_diffractive} we show the nuclear modification factor $R_{pA}$ for a charm quark with transverse momentum $\ptc$ at fixed photon energy $q^+=2p^+$, defined as
\begin{equation}\label{eq:defRpa}
  R_{pA} = \frac{\der\sigma^{\gamma A}}{A\, \der\sigma^{\gamma p}} \,.
\end{equation}

To understand the behavior of this ratio for the exclusive process in \fig\ref{fig:RpA_exclusive}, we first consider the dilute limit of the nuclear dipole amplitude \cite{Lappi:2013zma}: $\mathcal{N}_A(r_\perp,\bd)=T_A(\bd)\,S_\perp\,\mathcal{N}_p(r_\perp)$, with $T_A$ normalized to $A$. 
This estimate corresponds to what is known in this context as the Impulse Approximation (IA). 
Since the cross section is proportional to $\mathcal{N}^2$, this gives
\begin{equation}
  R_{pA}= \frac{S_\perp}{A}\int \der^2\bb_d\;T_A^2(\bd)\;\propto\;A^{1/3}\,,
  \label{eq:RpA_limit}
\end{equation}
which is $2.5$ for a lead nucleus, independent of $\ptc$. However, the $R_{pA}$ in \fig\ref{fig:RpA_exclusive} is below the IA, especially at small $\ptc$. This indicates that even for charm quarks, there is a significant contribution from dipoles that are large enough to be in the saturation regime $r_\perp \sim 1/\qs$.
In addition, the BK evolution reduces $\mathcal{N}_A/\mathcal{N}_p$ as $\xpo$ decreases: for small dipoles $\mathcal{N}\propto(r_\perp^2\qs^2)^{\gamma_\text{eff}}$ with an effective anomalous dimension $\gamma_\text{eff}<1$, so the ratio becomes $(Q_{s,A}^2/Q_{s,p}^2)^{\gamma_\text{eff}}$ \cite{Mueller:2002zm, Iancu:2002tr}. Moreover, with running coupling the nuclear saturation scale also grows more slowly \cite{Mueller:2003bz, Iancu:2007af, Albacete:2007yr}. As $\ptc$ increases, the dipoles become smaller and the nucleus becomes dilute, so $R_{pA}$ grows. This growth is, however, slow, because $\xpo\approx2m_Te^{-y}/\sqrt{s_{NN}}$ increases only linearly with $m_T$, and therefore an evolution effect remains. For the same reason, $R_{pA}$ is smaller at larger rapidity, where $\xpo$ is smaller. As a result, the limit \eq\eqref{eq:RpA_limit} is approached only at very large $\ptc$, where the relevant dipoles are small compared to $1/Q_{s,A}$ and, in addition, $\xpo>x_0=0.01$, so that no evolution effect remains.

\begin{figure}
    \centering
    \includegraphics[width=\linewidth]{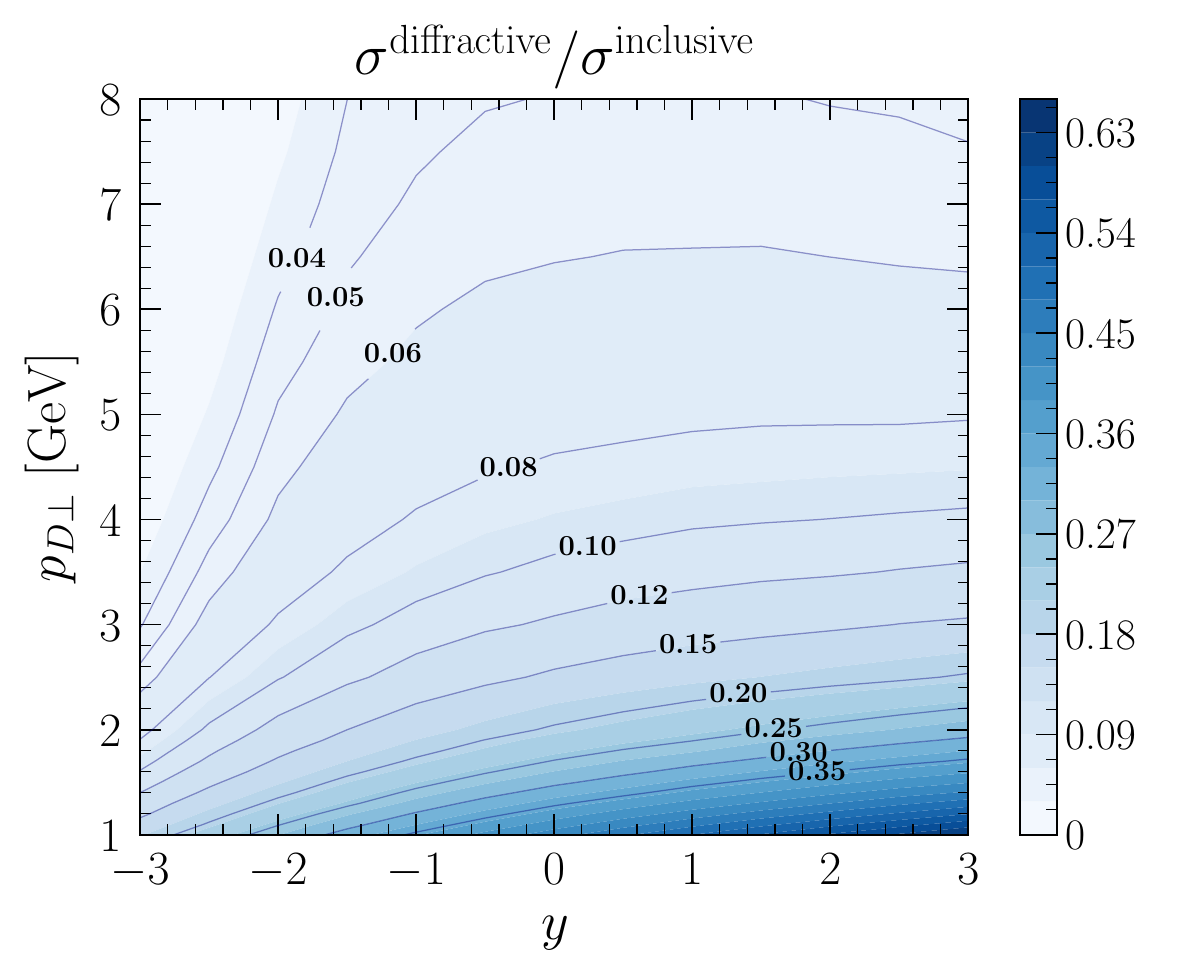}
    \caption{Ratio of the diffractive to the inclusive $D^0$ photoproduction cross section in Pb--Pb UPCs in the $An0n$ class, as a function of the rapidity $y$ and the transverse momentum $p_{D\perp}\geq1$~GeV of the meson. The BCFY fragmentation function is used.}
    \label{fig:ratio_An0n_PbPb}
\end{figure}

\begin{figure*}
    \centering
    \includegraphics[width=0.8\linewidth]{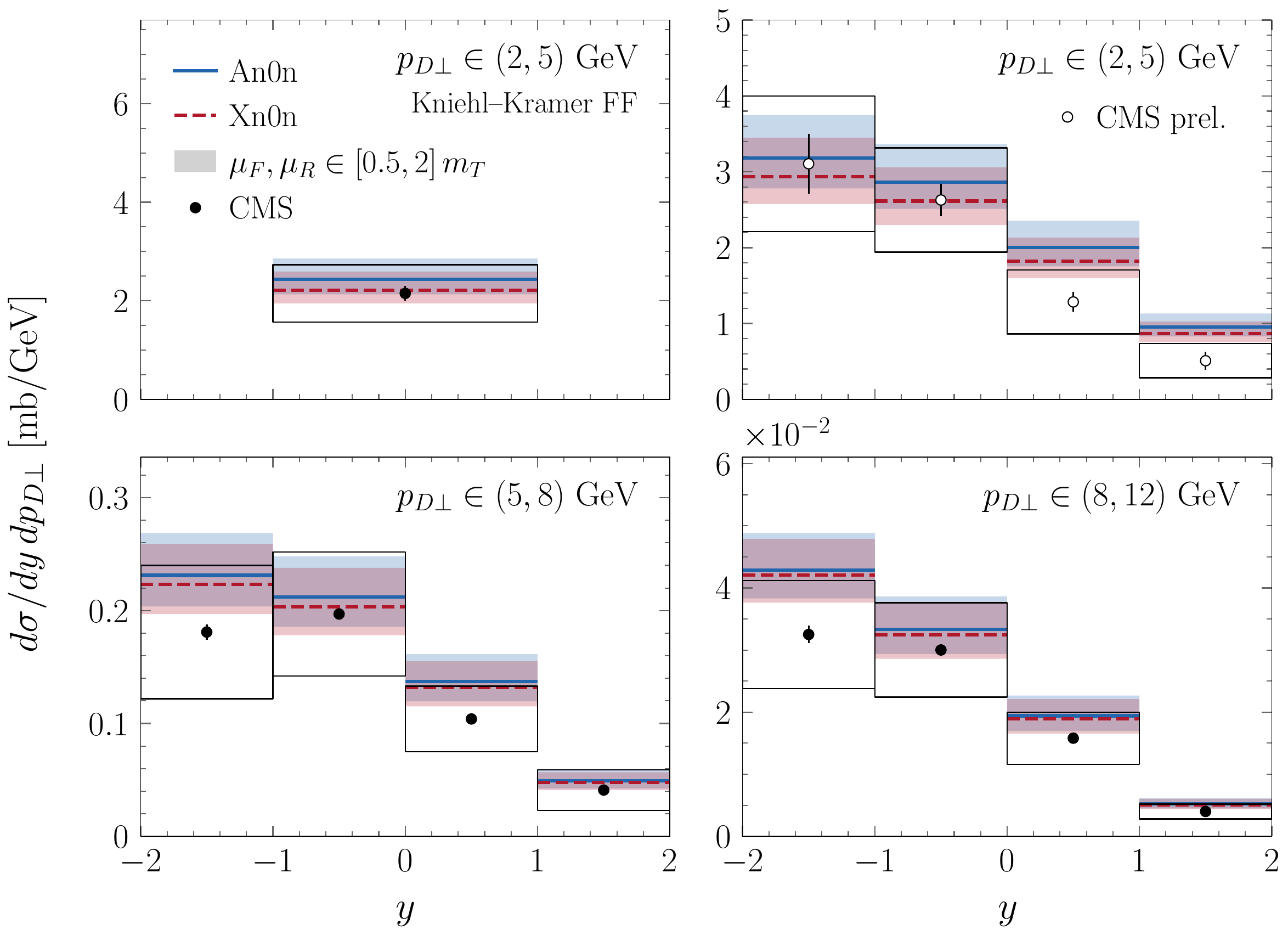}
    \caption{Inclusive $D^0$ photoproduction in Pb--Pb UPCs in the $An0n$ class (solid) and in the $Xn0n$ class (dashed), \eq\eqref{eq:inclusiveCrosssection}, with the Kniehl--Kramer fragmentation function. The CMS data~\cite{CMS:2025jjx} are the sum of the $Xn0n$ and rapidity-reflected $0nXn$ event classes, and we therefore multiply our predictions by a factor of 2. The $2<p_{D\perp}<5$~GeV bin is shown twice: in the top-left panel with the published data~\cite{CMS:2025jjx}, and in the top-right panel with the preliminary results of Ref.~\cite{CMS:2025tql} in four rapidity bins. Bars show the statistical and boxes the systematic uncertainties. The bands show the scale uncertainty, $\mu_F,\mu_R\in[0.5,2]\,m_T$.}
    \label{fig:D0_incl_corrected}
\end{figure*}

\begin{figure*}
    \centering
    \includegraphics[width=0.8\linewidth]{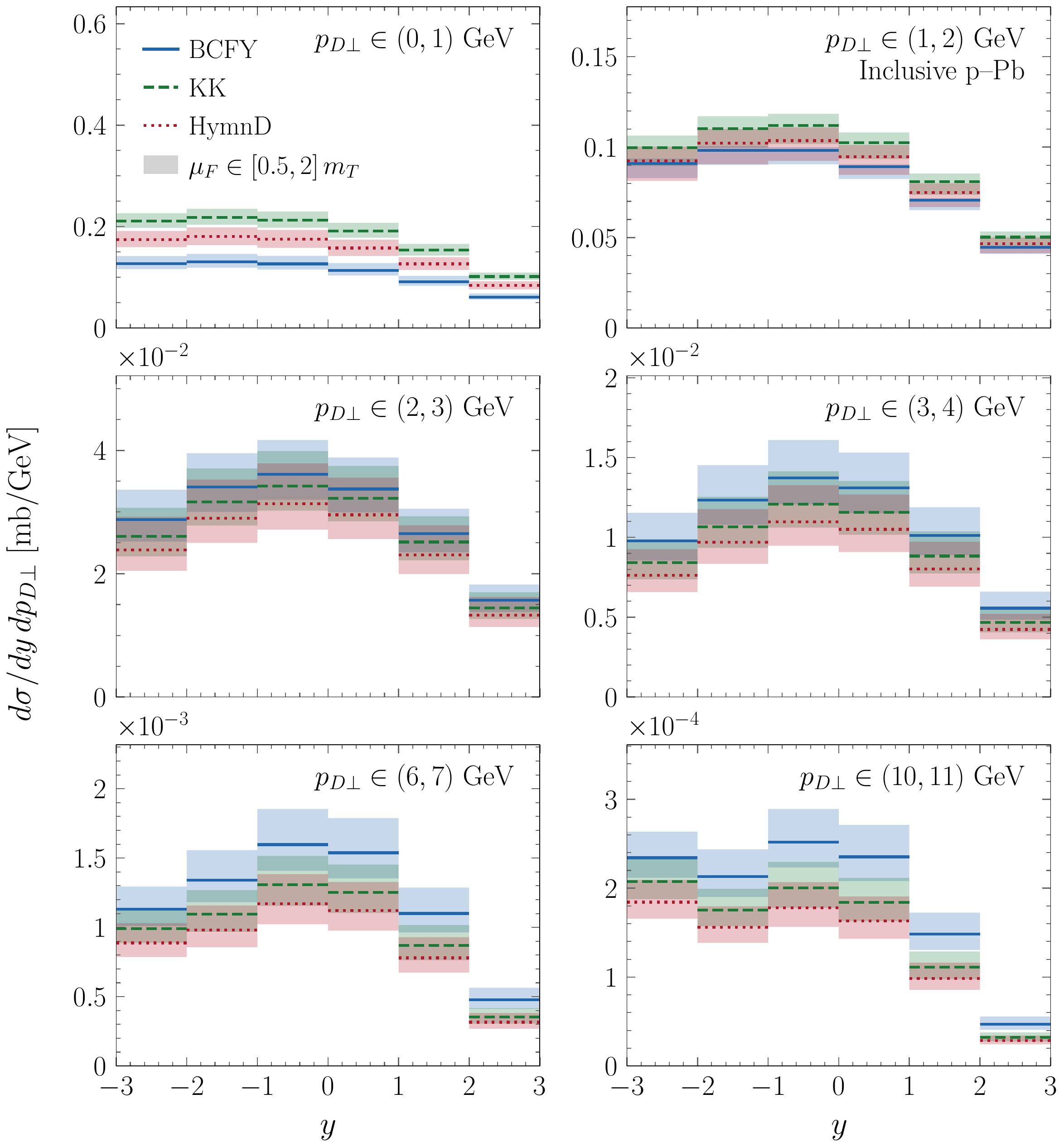}
    \caption{Inclusive $D^0$ photoproduction in p--Pb UPCs at $\sqrt{s_{NN}}=8.16$~TeV, with the photon emitted by the lead nucleus, for the BCFY (solid), KK (dashed) and HymnD (dotted) fragmentation functions. The bands show the scale variation uncertainty $\mu_F\in[0.5,2]\,m_T$.}
    \label{fig:D0_incl_bins_pPb}
\end{figure*}

On the other hand, the diffractive$\tmd$ contribution involves a large gluon--gluon dipole \eq\eqref{eq:G}, with $R_\perp\sim1/\qs$, which scatters closer to the Black Disk (BD)  limit \cite{Iancu:2021rup,Iancu:2022lcw}. In the BD limit, the gluon occupation number of the pomeron is the same for the proton and the nucleus \cite{Hauksson:2024bvv}, resulting in 
$G\to1-x$ in \eq\eqref{eq:G} for $K_\perp\ll \qs$ for both targets, i.e.\ $G\to1$ at small $x$. Thus in the BD limit the gluon DTMD \eq\eqref{eq:GDTMD} is simply proportional to the transverse area of the target, $S_\perp$ for the proton and approximately $\pi R^2$ for the nucleus. This alone would give $R_{pA}\approx\pi R^2/(AS_\perp)\propto A^{-1/3}$, which is $0.4$ for lead.  On the other hand the integration also extends to  $K_\perp\gg \qs$, where the gluon dipole is small and the nucleus is dilute. This regime gives a contribution  $G\propto \qs^2$ and $R_{pA}\propto A^{1/3}$ corresponding to the IA limit. The result in Fig.~\ref{fig:RpA_diffractive} reflects a delicate balance between the BD and IA regimes. 
We have  checked this interpretation with the dipole amplitudes used here, in the limit $x\to0$ of \eq\eqref{eq:G}: the $R_{pA}$ of the gluon DTMD is $0.42$ at $K_\perp\ll \qs$ and $1.3$--$2.5$ at $K_\perp\approx6$~GeV. Integrating it up to $K_\perp=3$--$20$~GeV gives $R_{pA}=0.7$--$0.96$, for $10^{-4}\leq\xpo\leq10^{-2}$. Overall, we emphasize that the approach of $R_{pA}$ to a value that is close to 1 at high $\ptc$ is not obvious, but reflects a balance of different behaviors in different regimes of $K_\perp$. 

Comparing Figs.~\ref{fig:RpA_exclusive} and~\ref{fig:RpA_diffractive} we see a stronger nuclear enhancement for the exclusive process than for the diffractive one. This reflects the  small size and consequently weaker interaction of the partonic quark-antiquark contribution in the former. This systematic  mirrors the nuclear modification results found for the diffractive structure function for massless quarks in Ref.~\cite{Kowalski:2008sa}. There the large-$\beta$ $q\bar{q}$ regime corresponds to our exclusive contribution, exhibiting a nuclear enhancement, and the 
small-$\beta$ $q\bar{q}g$ contribution to our diffractive regime,  with a nuclear suppression, when normalized with $A$ as in \eq\eqref{eq:defRpa}. 

\subsection{Inclusive \texorpdfstring{$D^0$}{D0} photoproduction in the \texorpdfstring{$Xn0n$}{Xn0n} class}
\label{sec:inclusive}
The predictions for inclusive $D^0$ production in Pb--Pb UPCs in Ref.\,\cite{Gimeno-Estivill:2025rbw} included the $0n$ condition on the photon-emitting nucleus, but not the $Xn$ condition on the target imposed in the experimental measurement. Using the results of the previous section, we can now estimate this correction. Following Ref.\,\cite{Guzey:2026lfb}, we write
\begin{equation}
\label{eq:inclusiveCrosssection}
    \der\sigma^{\text{inclusive}}_{Xn0n}=\der\sigma^{\text{inclusive}}_{An0n}-\der\sigma^{\text{diffractive}}_{0n0n} \,.
\end{equation}
Here, the first term is the inclusive cross section with the $0n$ condition on the photon-emitting side and no condition on the target, while the second term removes the coherent diffractive events, \eq\eqref{eq:diffractiveCS}, in which the target stays intact. In principle, the subtraction should include all $\xpo$ values. However, the target must absorb a minimal momentum transfer $|t_{\min}|\approx\xpo^2m_N^2$, and the nuclear form factor suppresses coherent scattering when $\sqrt{|t_{\min}|}$ exceeds the inverse nuclear radius, imposing the parametric constraint $\xpo\lesssim1/(m_NR)\approx0.03$ for coherent events. 
Quantitatively, only about $5\%$ of the coherent cross section lies at $x_{\mathbb{P}}>0.1$, meaning that approximately $95\%$ is concentrated in the region $x_{\mathbb{P}}<0.1$. The cut $\xpo<0.1$ therefore includes essentially the full coherent contribution. Note that the diffractive events where the target breaks up through an additional, independent EMD pass the $Xn$ selection and are therefore not subtracted.

For both terms, we use the photon fluxes of Sec.\,\ref{sec:PhotonFlux} and the fragmentation functions of Sec.\,\ref{sec:FF}, while the CGC calculation of the inclusive cross section is described in Ref.\,\cite{Gimeno-Estivill:2025rbw}.

First, in \fig\ref{fig:ratio_An0n_PbPb} we show the ratio of the diffractive to the inclusive $D^0$ cross section in Pb--Pb UPCs in the $An0n$ class. We show it only for $p_{D\perp}\geq1$~GeV, since at lower momentum the DTMD approach, derived in the limit where there is a clear hard scale in the problem, is less reliable. Note also that our inclusive cross section is computed at leading order, while the coherent $q\bar qg$ contribution is formally part of its NLO correction, so that \eq\eqref{eq:inclusiveCrosssection} subtracts a term which the leading-order calculation does not contain. We nevertheless use it as our best estimate of the coherent events removed by the $Xn$ condition. The ratio is largest at the smallest $p_{D\perp}$ and forward rapidity, where both calculations are only estimates (see Sec.~\ref{sec:diffractiveTMD}), and it decreases with $p_{D\perp}$, because the diffractive spectrum falls faster than the inclusive cross section. At fixed $p_{D\perp}$, the ratio grows towards positive rapidity, where $\xpo$ is smaller and the BK evolution is longer. Conversely, at negative rapidity and large $p_{D\perp}$ the ratio is small, because $\xpo$ approaches the limit $\xpo\rightarrow0.1$. Overall, this shows that the diffractive contribution is important for the inclusive measurement mainly at small $p_{D\perp}$ and forward rapidity.

A similar ratio was computed in Ref.\,\cite{Guzey:2026lfb} with LTA nuclear diffractive PDFs. The dependence on $p_{D\perp}$ and $y$ is the same, but at small $p_{D\perp}$ our ratio is larger than in both their ``low'' and ``high'' shadowing scenarios, whereas at large $p_{D\perp}$ both calculations give a few percent. Since the ratio also depends on the inclusive cross section in the denominator, part of this difference can come from the different inclusive calculations: the CGC framework of Ref.\,\cite{Gimeno-Estivill:2025rbw} here, and the G$\gamma$A--FONLL framework with collinear nuclear PDFs in Ref.\,\cite{Guzey:2026lfb}. Moreover, at $p_{D\perp}\approx2$~GeV our ratio, $0.1$--$0.25$, lies between their LTA results and their impulse-approximation estimate, $\sim30\%$, which they consider an upper bound. Our ratio exceeds this estimate only at $p_{D\perp} \lesssim 1.5$~GeV and forward rapidity, where, however, both our diffractive$_{\rm TMD}$ and our leading-order inclusive results are less reliable.

Next, in \fig\ref{fig:D0_incl_corrected} we compare the inclusive $D^0$ cross section in the $An0n$ class with the $Xn0n$ prediction of \eq\eqref{eq:inclusiveCrosssection} and with the CMS data \cite{CMS:2025jjx,Ogrodmk:2026wqz,CMS:2025tql}. Here, and only in this figure, we multiply our cross section by a factor 2, to account for the fact that the experimental data is reported as a  ``sum of the cross section measured in $Xn0n$ events, and the $y$-reflected cross section measured in $0nXn$ events.'' Additionally, we make the approximation that the $D^0$ and $\bar{D}^0$ cross sections are equal. We use here the KK fragmentation function, which results from a LO fit to experimental data, consistently with our LO hard cross section \footnote{Note that the CMS data include both prompt and nonprompt $D^0$ mesons~\cite{CMS:2025jjx}, while our calculation includes only the $c\to D^0$ channel. The $b\to D^0$ contribution was estimated in Ref.~\cite{Gimeno-Estivill:2025rbw} to be small at low $p_{D\perp}$ and to grow with $p_{D\perp}$.}. The correction from the diffractive cross section is about $10\%$ for $p_{D\perp}\in[2,5]$ GeV, decreases at larger momentum and depends only weakly on rapidity. 
For $2<p_{D\perp}<5$ GeV, our prediction is in a rather good agreement with data, while developing a tendency to overestimate it, although still within the error bands, for higher $p_{D\perp}$. This is in line with our expectation that the CGC calculation should be most reliable for semihard momenta while, as discussed in Sec.~\ref{sec:diffractiveTMD}, an additional DGLAP evolution of the small-$x$ gluon PDF (and the DPDFs in the coherent diffractive subtraction term) should be included for higher $p_{D\perp}$. In fact the quantitatively larger difference to our earlier calculation of inclusive cross sections in Ref.~\cite{Gimeno-Estivill:2025rbw} is that we now include the DGLAP evolution of the FF, which was earlier taken at a fixed scale.

Finally, \fig\ref{fig:D0_incl_bins_pPb} shows the predictions for inclusive $D^0$ production in p--Pb UPCs. Note again that the rapidity range in this plot extends down to $y = -3$, where the momentum fraction of the target probed by the dipole, $x = m_T e^{-y}/\sqrt{s_{NN}}$ with $m_T$ the transverse mass of the charm quark, may no longer be sufficiently small for the CGC formalism to remain applicable. In the largest $p_{D\perp}$ bin, $x$ exceeds $x_0 = 0.01$ for $y \lesssim -2$, where the dipole amplitude is frozen at its initial condition. This causes the rise of the cross section towards $y = -3$ in that bin.
The dependence on the fragmentation function is the same as in the diffractive production, \fig\ref{fig:D0_bins_pPb}. At $p_{D\perp}<2$ GeV, the cross section follows the normalization of the FFs, with KK the largest and BCFY the smallest. At $2<p_{D\perp}<3$ GeV, the three predictions are close, while at higher $p_{D\perp}$ BCFY gives the largest cross section.
\section{Summary}
\label{sec: summary}
We have presented predictions for coherent diffractive $D^0$ photoproduction in Pb--Pb and p--Pb UPCs at the LHC in the Color Glass Condensate framework, including the exclusive production of a quark-antiquark pair and the production of a quark-antiquark-gluon system in diffractive TMD factorization. Although formally suppressed by $\as$, the latter dominates, since the recoil is taken by a semi-hard gluon that scatters strongly off the target, while the exclusive channel is suppressed by color transparency at large $\ptc$. The diffractive cross section is largest at backward and central rapidity, depends on the upper limit $x_{\mathbb{P}\max}$, and is only an estimate at $\ptc\lesssim2$~GeV. Compared with collinear calculations based on diffractive PDFs, our $p_{D\perp}$ spectrum is harder. The nuclear modification factor is close to one for the diffractive$\tmd$ channel, while for the exclusive channel it grows with transverse momentum up to about a factor of 2, below the dilute limit.

We also used the diffractive cross section to correct the inclusive $D^0$ cross section for the $Xn0n$ selection of the CMS measurement. This correction is roughly $10\%$ for $2<p_{D\perp}<5$~GeV and smaller at higher momenta. Our calculation of inclusive $D$-meson production is now in better agreement with the CMS measurement than the one reported in Ref.~\cite{Gimeno-Estivill:2025rbw}, partially because of this correction but more importantly due to the inclusion of DGLAP evolution of the fragmentation function. The uncertainty related to hadronization  and the scale variation uncertainty that we have demonstrated here, remain as important factors limiting the accuracy of predictions for experimental data.  Finally, we presented predictions for inclusive $D^0$ photoproduction in p--Pb UPCs, which we hope will be examined in future analyses by the LHC experiments.

There are several avenues for improvement in the calculation that would be interesting to pursue in the future. For the diffractive process, at lower transverse momenta it would be important to also include the virtual corrections, to achieve a full NLO accuracy for the cross section. At high transverse momenta, further investigations are also warranted into the effects of DGLAP evolution of the quark and gluon DPDFs, which captures the leading NLO corrections to the diffractive process (which are NNLO corrections to the exclusive one) beyond those enhanced by $\ln(1/\xpo)$ already included in our calculation. Moreover, the inclusive cross section calculation used here is only computed at leading order.  In the future one should also include real and virtual gluon radiation, bringing it to the same NLO level as the diffractive one. This would require extending the NLO results~\cite{Bergabo:2022zhe,Caucal:2024cdq,Altinoluk:2025dwd} for single-inclusive hadroproduction in DIS with massless quarks to the massive-quark case, building on recent progress in light-cone perturbation theory with massive partons~\cite{Beuf:2021srj,Beuf:2022ndu}. 

To probe the genuine quark and gluon diffractive TMDs, one should also consider diffractive back-to-back production of a pair of $D$ (or $B$) mesons or of a $D$-meson-tagged jet in association with a light jet, measuring both the relative and total transverse momenta of the pair. This would require accounting for the Sudakov suppression of these distributions~\cite{Mueller:2012uf,Mueller:2013wwa,Hatta:2021jcd,Iancu:2025jsu}, including mass effects, following the approaches developed in~\cite{Gao:2026azd,Salazar:2026bmt} for the analogous inclusive process.
Another important channel for heavy quark physics in UPCs, both inclusive and diffractive, is quarkonium production. Here, for small transverse momentum one could use the NRQCD approach~\cite{Kang:2013hta,Gimeno-Estivill:2024gbu,Cheung:2024qvw} to the hadronization of the quark-antiquark pair into a $J/\Psi$ or $\Upsilon$ vector meson.

\begin{acknowledgments}
We thank Carlisle Casuga, Vadim Guzey, Felix Hekhorn, Ilkka Helenius, Gian Michele Innocenti, Anna-Mari Levälampi, Heikki Mäntysaari, Petja Paakkinen and Anna M. Staśto for useful discussions. 
This work was supported by the Research Council of Finland, the Centre of Excellence in Quark Matter (projects 346324 and 364191), and by the European Research Council (ERC, grant agreements  No. ERC-2023-101123801 GlueSatLight and ERC-2018-ADG-835105 YoctoLHC).
Computing resources from CSC – IT Center for Science in Espoo, Finland and the Finnish Grid and Cloud Infrastructure (persistent identifier \texttt{urn:nbn:fi:research-infras-2016072533}) were used in this work. P.G.E. acknowledges  support from the Finnish Cultural Foundation. P.C.\ is funded by the Agence Nationale de la Recherche under
grant ANR-25-CE31-5230 (TMD-SAT) and by the CNRS/IN2P3 Master Project QHAT.
F.S.\ is supported by the Laboratory Directed Research and Development of Brookhaven National Laboratory and RIKEN-BNL Research Center, as well as the National Science Foundation (NSF) within the framework of the JETSCAPE collaboration, under grant number OAC-2514008 (CSSI:C-SCAPE). P.C.\ and F.S.\ also acknowledge the Saturated Glue (SURGE) Topical Theory Collaboration, funded by the U.S.\ Department of Energy, Office of Science, Office of Nuclear Physics. The content of this article does not reflect the official opinion of the European Union and responsibility for the information and views expressed therein lies entirely with the authors. 

\end{acknowledgments}
\FloatBarrier

\appendix
\section{Photon flux: target-size effects and STARlight comparison}
\label{sec:flux_approximations}
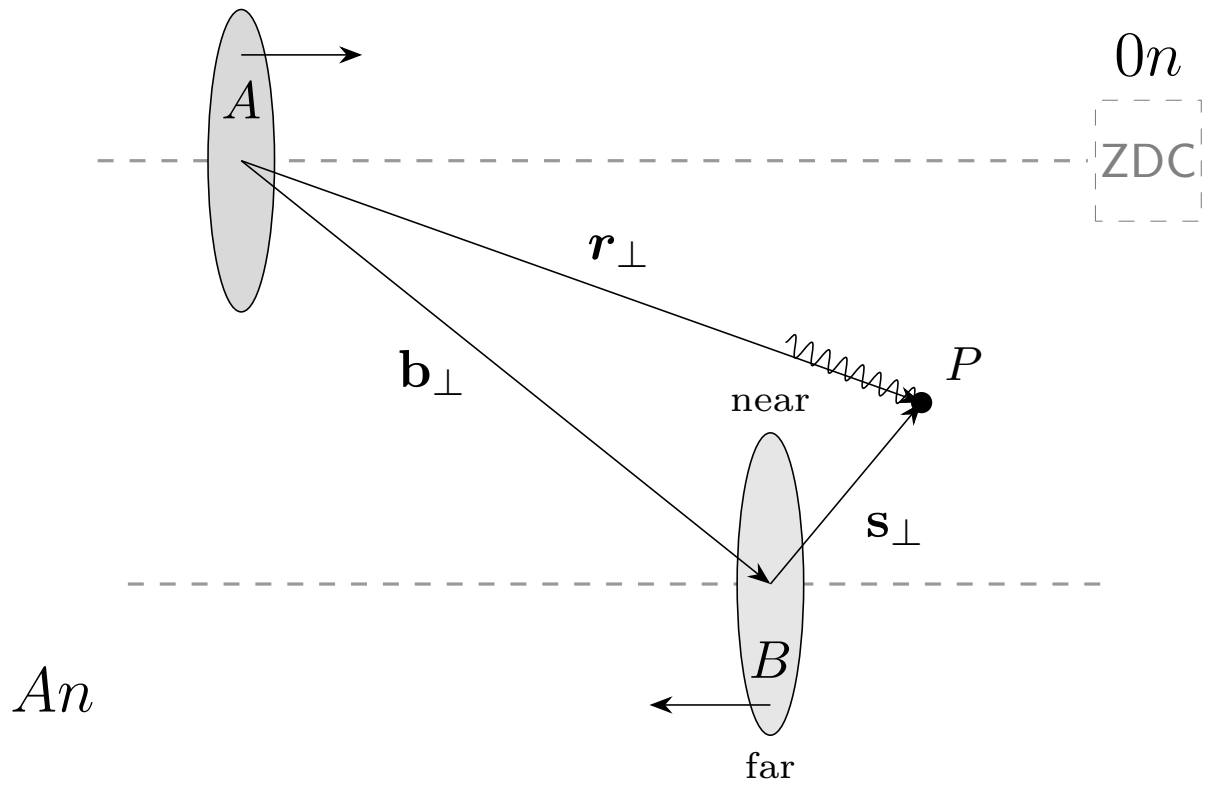
\begin{figure}[tbh!]
    \centering
    \begin{tikzpicture}[
  >=Stealth,
  every node/.style={font=\normalsize},
  vec/.style={->, line width=0.3pt},
  photon/.style={decorate, decoration={snake, amplitude=2pt, segment length=3.5pt, pre length=0mm, post length=0mm}, line width=0.3pt},
]
\coordinate (A) at (0, 0);
\coordinate (B) at (3.5, -2.8);
\coordinate (P) at (4.5, -1.6);
 
\draw[dashed, gray!80, line width=0.6pt] (-0.95, 0) -- (5.6, 0);
\draw[dashed, gray!80, line width=0.6pt] (-0.75, -2.8) -- (5.75, -2.8);
 
\draw[fill=gray!30, draw=black, line width=0.3pt] (A) ellipse (0.22 and 1.0);
\draw[->, line width=0.3pt] (0, 0.7) -- (0.8, 0.7);
 
\draw[fill=gray!20, draw=black, line width=0.3pt] (B) ellipse (0.22 and 1.0);
\draw[->, line width=0.3pt] (3.5, -3.6) -- (2.7, -3.6);

\node[font=\scriptsize] at ($(B)+(0, 1.2)$) {near};
\node[font=\scriptsize] at ($(B)+(0, -1.2)$) {far};
\node[] at ($(A)+(0, 0.4)$) {$A$};
\node[] at ($(B)+(0, -0.5)$) {$B$};
\fill (P) circle (2pt);
\node[above right=1pt, font=\small] at (P) {$P$};
 
\draw[vec] (A) -- (P);
\node at (2.5, -0.6) {$\boldsymbol{r}_\perp$};
\draw[photon] (3.6,-1.2) to (P);

\draw[vec] (A) -- (B) node[midway, left=4pt] {$\bb_\perp$};
\draw[vec] (B) -- (P) node[midway, below right=0.5pt] {$\sbo_\perp$};
\node[font=\large] at (-1.25, -3.5) {$An$};
\draw[gray, dashed, line width=0.2pt] (5.65, -0.4) rectangle (6.35, 0.4);
\node[font=\small, gray] at (6.0, 0.0) {\textsf{ZDC}};
\node[font=\large] at (6.0, 0.7) {$0n$};
\end{tikzpicture}
    \caption{Transverse-plane geometry of a UPC in the $An0n$ class. The  nucleus $A$ emits a quasi-real photon (wavy line) which is absorbed by a nucleon at point $P$ inside the target nucleus $B$. The vector $\rb_\perp$ is the distance from the center of the nucleus $A$ to the interaction point $P$, $\bb_\perp$ is the impact parameter between the centers of the two nuclei, and $\sbo_\perp = \rb_\perp - \bb_\perp$ is the position of $P$ measured from the center of the target. The labels ``near'' and ``far'' indicate the nucleons in the target closest and farthest from the nucleus $A$, respectively. Experimentally, the ZDC on the photon-emitting side detects no forward neutrons ($0n$), while the $An$ event class corresponds to no requirement imposed on the target side.}
    \label{fig:UPC}
\end{figure}
\begin{figure}[tbh!]
    \centering
    \includegraphics[width=1\linewidth]{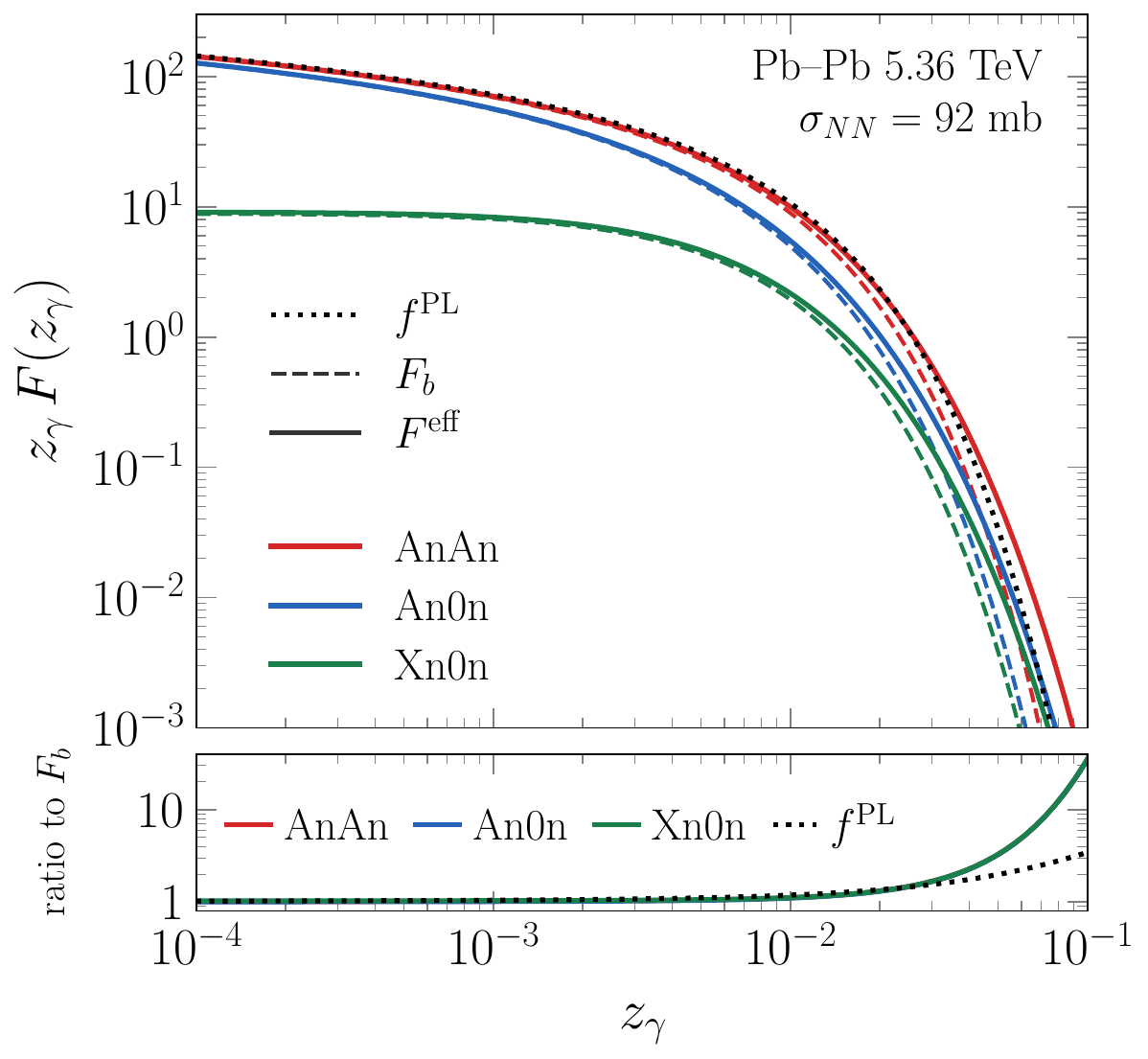}
    \caption{Photon flux $z_\gamma F(z_\gamma)$ in Pb--Pb collisions at $\sqrt{s_{NN}}=5.36$ TeV as a
    function of the photon energy fraction $z_\gamma$, for the neutron classes $AnAn$, $An0n$ and
    $Xn0n$ in different colors. In dashed lines, the photon flux \eq\eqref{eq:flux_b} is evaluated at the center of the target nucleus, at a distance
    $|\mathbf b|$ from the photon-emitter. In solid lines, the effective flux \eq\eqref{eq:eff_photon_flux} is evaluated
    at the interaction point $P$, at a distance $|\mathbf r|$ from the emitter, and averaged over its
    position $\mathbf s$ in the target (see \fig\ref{fig:UPC}). With a dotted line we represent the
    point-like approximation \eqref{eq:pointlikePhotonFlux} for $AnAn$ with $b_\perp>b_{\min}=14.2$ fm. Both $F_b$ and
    $F^{\mathrm{eff}}$ use $\sigma_{NN}=92$ mb. The ratios to $F_b$ for different neutron emission classes  in the lower panel overlap so closely that they are not visible. }
    \label{fig:WS_vs_eff}
\end{figure}
\begin{figure}[tbh!]
    \centering
    \includegraphics[width=1\linewidth]{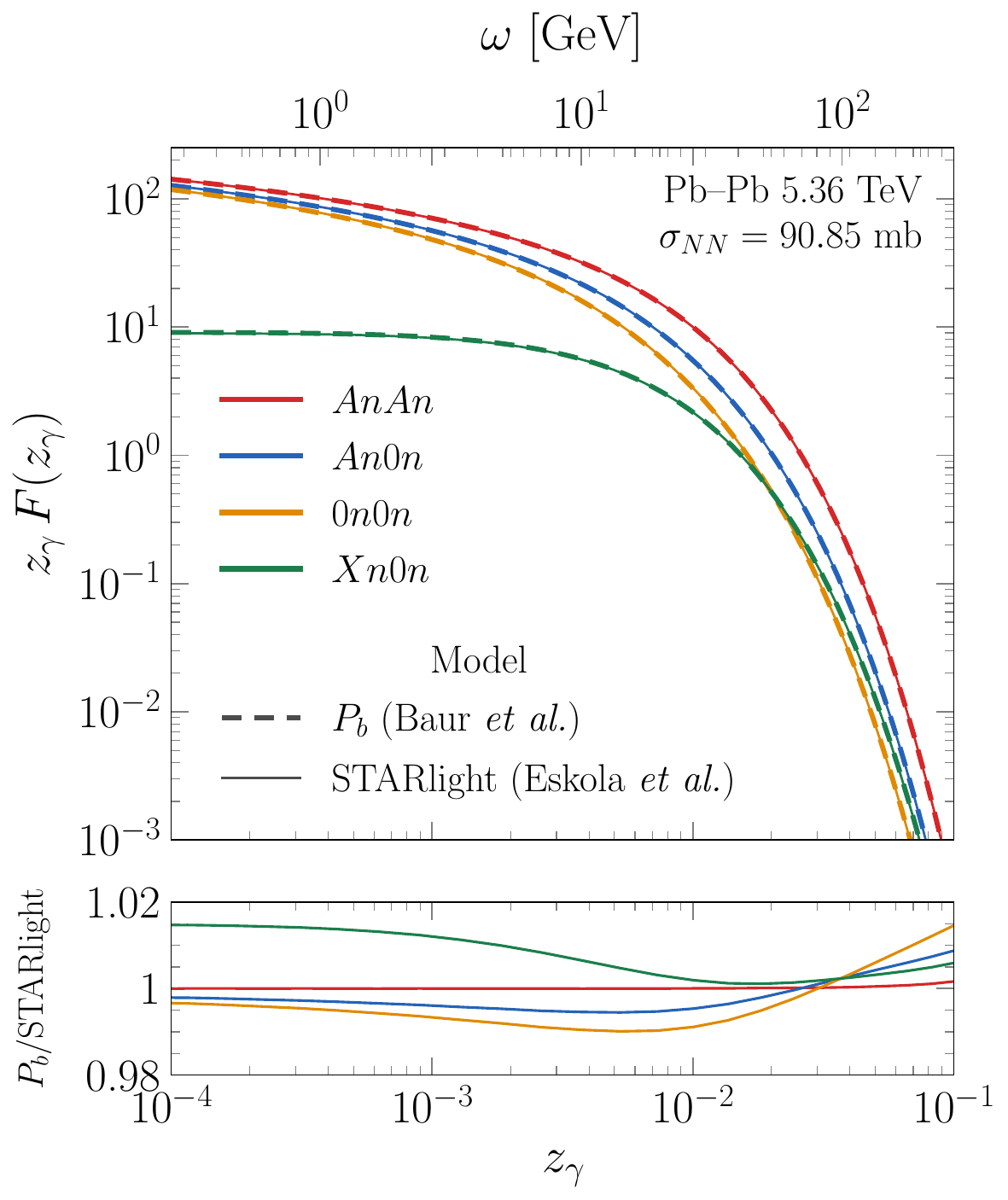}
    \caption{Effective photon flux $z_\gamma F(z_\gamma)$ in Pb--Pb collisions at $\sqrt{s_{NN}}=5.36$~TeV as a function of the photon energy fraction $z_\gamma$ (bottom axis) and photon energy $\omega$ (top axis), for different neutron classes (colors). The dashed lines correspond to the calculation with the electromagnetic-dissociation probability $P(b_\perp)$ of
\eq\eqref{eq:EMD}. The thin solid lines correspond to the photon flux tables of Ref.\,\cite{Eskola:2024fhf}, computed with
STARlight~\cite{Klein:2016yzr}. The tables do not include the $Xn0n$ class, which we obtain from them as $An0n-0n0n$. Both use $\sigma_{NN}=90.85$ mb \cite{Eskola:2024fhf}. The lower panel shows the ratio of our photon flux to STARlight.}
    \label{fig:Starlight_vs_Pb}
\end{figure}
\begin{figure}[tbh!]
    \centering
    \includegraphics[width=1\linewidth]{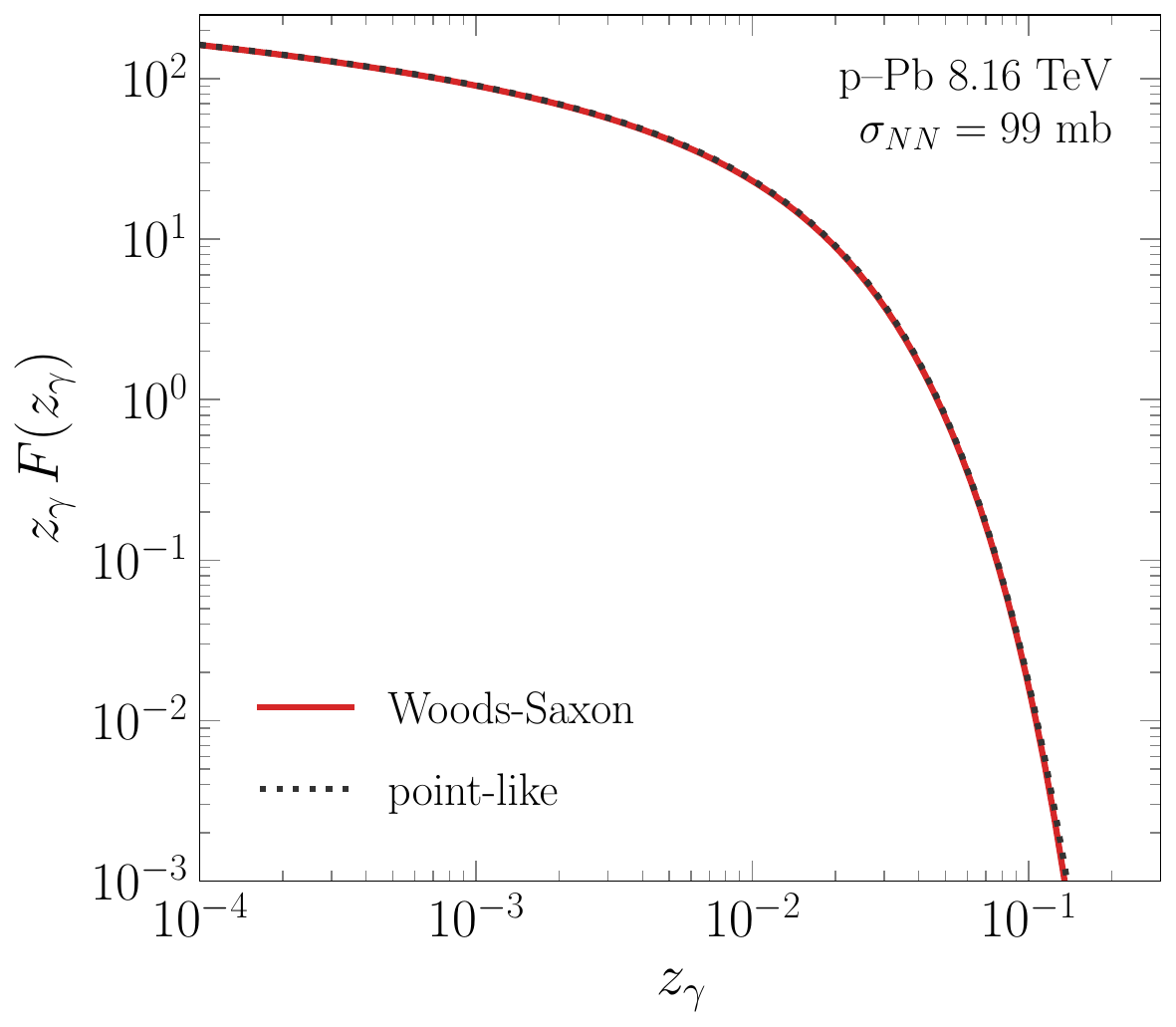}
    \caption{Photon flux $z_\gamma F(z_\gamma)$ of the lead nucleus in p--Pb collisions at
$\sqrt{s_{NN}}=8.16$~TeV as a function of the photon energy fraction $z_\gamma$. The solid-red line is the photon flux from a Woods--Saxon charge distribution with the survival factor in \eq\eqref{eq:pA_survival_factor}, with $\sigma_{NN}=99$~mb. The dotted line corresponds to the point-like approximation \eqref{eq:pointlikePhotonFlux} with a
sharp cut $b_\perp>b_{\min}=1.1R$, with $R=7.1$ fm. No electromagnetic-dissociation factor is
applied.}
    \label{fig:proton_flux}
\end{figure}

In this appendix, we quantify the difference between the flux used in Sec.~\ref{sec:PhotonFlux}
and a simplified version that neglects the transverse size of the target. We then compare two
parametrizations of the electromagnetic-dissociation probability, and the flux in proton--lead
collisions with its point-like approximation.

In \eq\eqref{eq:eff_photon_flux}, we evaluate the photon flux $F^{\text{eff}}$ at the transverse position $\rb$ of the interaction point $P$, i.e., at the separation between the center of the photon-emitting nucleus and the struck nucleon (\fig\ref{fig:UPC}). A common simplification is to evaluate the photon flux at the impact parameter $\bb$, which neglects the transverse extent of the target. Changing variables to $\bb=\rb-\sbo$ and evaluating the flux at $b_\perp$ instead of $r_\perp=|\bb+\sbo|$, the $\sbo$ integral reduces to $\int\der^2\sbo\,T_B(\sbo)/B=1$, and
\begin{equation}
\label{eq:flux_b}
F_b(\zp)=\int\der^2\bb\, f_{\gamma/A}^{\text{WS}}(\zp,b_\perp)\,\Gamma_{AB}(b_\perp)\,P_{\text{EMD}}^{\text{channel}}(b_\perp)\,.
\end{equation}

As shown in \fig\ref{fig:WS_vs_eff}, evaluating the flux $F_b$ at the center of the target
underestimates it at large $\zp$, by a factor of about $2$ at $\zp=0.03$ and about $30$ at
$\zp=0.1$, while both fluxes agree within about $10\%$ for $\zp\leq10^{-2}$.
This follows from the point-like flux \cite{Baur:2001jj, Bertulani:2005ru, Baur:1998ay}, which is the same as the Woods--Saxon flux \eqref{eq:WS_bare_photon_flux} outside the nucleus,
\begin{equation}
   \label{eq:pointlikePhotonFlux}
     f^{\text{PL}}_{\gamma/A}(\zp, x_\perp) = \frac{\aem Z^2}{\pi^2}\frac{1}{\zp}\big(\zp m_N\operatorname{K}_1(\zp m_N x_\perp)\big)^2\,,
\end{equation}
where $x_\perp$ is the transverse distance from the emitter. For $\zp m_N x_\perp\gg1$ the Bessel function behaves as $e^{-2\zp m_N x_\perp}/x_\perp$, so it falls off on the scale $1/(\zp m_N)$. At large $\zp$ the argument of the Bessel function varies strongly across the target, and the $\bb$ integral is dominated by $b_\perp\approx2R$, where the near side of the target sees a flux larger than at the center by $\sim e^{2\zp m_N R}$, a factor of several hundred at $\zp=0.1$. At small $\zp$, $1/(\zp m_N)\gg R$, the flux is nearly constant across the target, and $F^{\text{eff}}$ and $F_b$ coincide. Consistently, the two curves in \fig\ref{fig:WS_vs_eff} separate around $\zp\sim10^{-2}$, where $1/(\zp m_N)\approx20$~fm becomes comparable to the size of the nucleus. This only affects the cross section at large transverse momentum and forward rapidity.

In \fig\ref{fig:Starlight_vs_Pb} we compare the effective flux computed with the EMD probability $P(b_\perp)$ in \eq\eqref{eq:EMD} with the STARlight EMD probability \cite{Klein:2016yzr, Eskola:2024fhf}. For this comparison, we use the same $\sigma_{NN} = 90.85$~mb as the STARlight flux. 

For the $AnAn$ channel without neutron selection, the two calculations agree throughout the $\zp$ range, which validates our implementation of the effective
flux. With neutron selection, using the $P(b_\perp)$ instead of the STARlight parametrization changes the flux $An0n$ and $0n0n$ by less than $1.5\%$.  The choice of the electromagnetic-dissociation probability is therefore negligible for the predictions.

Finally, \fig\ref{fig:proton_flux} shows the photon flux of the lead nucleus in p--Pb collisions
at $\sqrt{s_{NN}}=8.16$~TeV. This is computed from \eq\eqref{eq:pAphotonFlux} with $\sigma_{NN}=99$~mb and
no electromagnetic-dissociation factor. The figure also shows the point-like flux
\eq\eqref{eq:pointlikePhotonFlux} with a sharp cut $b_{\min}=1.1R$, with $R=7.1$ fm. The two agree within
$1.5\%$ for $\zp\leq10^{-2}$ and within $6\%$ at $\zp=0.1$, so in p--Pb collisions the flux is
insensitive to the treatment of small impact parameters, in agreement with
Ref.\,\cite{Guzey:2026lfb}.
\FloatBarrier

\bibliographystyle{JHEP-2modlong}
\bibliography{refs}
\end{document}